\def\tr{{\text{tr}}\,}

\def\sgn{{\text{sgn\,}}}
\def\be{\begin{equation}}
\def\ee{\end{equation}}
\def\bea{\begin{eqnarray}}
\def\eea{\end{eqnarray}}
\def\bse{\begin{subequations}}
\def\ese{\end{subequations}}

\documentclass[prb,twocolumn,showpacs,amsmath,amssymb,eqsecnum,preprintnumbers]{revtex4}
\usepackage{graphicx}% Include figure files
\usepackage{dcolumn}% Align table columns on decimal point
\usepackage{bm}% bold math
\begin{document}
\preprint{KITP Preprint NSF-KITP-05-85}
%\preprint{Phys. Rev. B {\bf xx}, xxxxxx (200x)}
\title{Theory of helimagnons in itinerant quantum systems}
\author{D. Belitz$^{1,2}$, T. R. Kirkpatrick$^{1,3}$, and A. Rosch$^4$}
\affiliation{$^{1}$ Kavli Institute for Theoretical Physics, University of
                    California, Santa Barbara, CA 93106, USA\\
             $^{2}$ Department of Physics and Materials Science Institute, University
                    of Oregon, Eugene, OR 97403, USA\\
             $^{3}$ Institute for Physical Science and Technology,
                    and Department of Physics, University of Maryland, College Park,
                    MD 20742, USA\\
             $^{4}$ Institut f{\"u}r Theoretische Physik, Universit{\"a}t zu K{\"o}ln,
                    Z{\"u}lpicher Strasse 77, D-50937 K{\"o}ln, Germany}
\date{\today}

\begin{abstract}
The nature and effects of the Goldstone mode in the ordered phase of helical or
chiral itinerant magnets such as MnSi are investigated theoretically. It is
shown that the Goldstone mode, or helimagnon, is a propagating mode with a
highly anisotropic dispersion relation, in analogy to the Goldstone mode in
chiral liquid crystals. Starting from a microscopic theory, a comprehensive
effective theory is developed that allows for an explicit description of the
helically ordered phase, including the helimagnons, for both classical and
quantum helimagnets. The directly observable dynamical spin susceptibility,
which reflects the properties of the helimagnon, is calculated.
\end{abstract}


\maketitle

\section{Introduction}
\label{sec:I}

Ferromagnetism and antiferromagnetism are the most common and well-known
examples of long-range magnetic order in solids. The metallic ferromagnets Fe
and Ni in particular are among the most important and well-studied magnetic
materials. In the ordered phase, where the rotational symmetry in spin space is
spontaneously broken, one finds soft modes in accord with Goldstone's theorem,
namely, ferromagnetic magnons. The latter are propagating modes with a
dispersion relation, or frequency-wave vector relation, $\Omega \sim {\bm k}^2$
in the long-wavelength limit. In antiferromagnets, the corresponding
antiferromagnetic magnons have a dispersion relation $\Omega \sim \vert{\bm
k}\vert$. In rotationally invariant models that ignore the spin-orbit coupling
of the electronic spin to the underlying lattice structure these relations hold
to arbitrarily small frequencies $\Omega$ and wave vectors ${\bm k}$. The
lattice structure ultimately breaks the rotational symmetry and gives the
Goldstone modes a mass. In ferromagnets, the low-energy dispersion relation is
also modified by the induced magnetic field, which generates a domain
structure. These are very small effects, however, and magnons that are soft for
all practical purposes are clearly observed, directly via neutron scattering,
and indirectly via their contribution to, e.g., the specific
heat.\cite{Kittel_1996_ch_15} These observations illustrate important concepts
of symmetries in systems with many degrees of freedom with ramifications that
go far beyond the realm of solid-state physics.\cite{Forster_1975,
Zinn-Justin_1996}

In systems where the lattice lacks inversion symmetry additional effects occur
that are independent of the spin rotational symmetry. This is due to terms in
the action that are invariant under simultaneous rotations of real space and
${\bm M}$, with ${\bm M}$ the magnetic order parameter, but break spatial
inversion symmetry. Microscopically, such terms arise from the spin-orbit
interaction, and their precise functional form depends on the lattice
structure. One important class of such terms, which is realized in the metallic
compound MnSi, is of the form ${\bm M}\cdot({\bm \nabla}\times{\bm
M})$.\cite{Dzyaloshinski_1958, Moriya_1960} They are known to lead to helical
or spiral order in the ground state, where the magnetization is
ferromagnetically ordered in the planes perpendicular to some direction ${\bm
q}$, with a helical modulation of wavelength $2\pi/\vert{\bm q}\vert$ along the
${\bm q}$ axis.\cite{Bak_Jensen_1980, Nakanishi_et_al_1980} In MnSi, which
displays helical order below a temperature $T_c \approx 30\,{\text K}$ at
ambient pressure, $2\pi/\vert{\bm q}\vert\approx 180\,{\text
\AA}$.\cite{Ishikawa_et_al_1976} Application of hydrostatic pressure $p$
suppresses $T_c$, which goes to zero at a critical pressure $p_{\,c} \approx
14\,{\text{kbar}}$.\cite{Pfleiderer_et_al_1997}

In addition to the helical order, which is well understood, MnSi shows many
strange properties that have attracted much attention lately and so far lack
explanations. Arguably the most prominent of these features is a pronounced
non-Fermi-liquid behavior of the resistivity in the disordered phase at low
temperatures for $p > p_{\,c}$.\cite{Pfleiderer_Julian_Lonzarich_2001} In part
of the region where non-Fermi-liquid behavior is observed, neutron scattering
shows `partial' magnetic order where helices still exist on intermediate length
scales but have lost their long-range directional
order.\cite{Pfleiderer_et_al_2004} Such non-Fermi-liquid behavior is not
observed in other low-temperature magnets. Since the helical order is the only
obvious feature that sets MnSi apart from these other materials it is natural
to speculate that there is some connection between the helical order and the
transport anomalies. In this context it is surprising that some basic
properties and effects of the helically ordered state, and in particular of the
helical Goldstone mode, which we will refer to as a {\em helimagnon} in analogy
to the ferromagnons and antiferromagnons mentioned above, are not known. The
purpose of the present paper is to address this issue. We will identify the
helimagnon and determine its properties, in particular its dispersion relation
and damping properties. We also calculate the spin susceptibility, which is
directly observable and simply related to the helimagnon. The effects of this
soft mode on various other observables will be explored in a separate
paper.\cite{paper_II} A brief account of some of our results, as well as some
of their consequences, has been given in Ref.\
\onlinecite{Kirkpatrick_Belitz_2005}.

One of our goals is to develop an effective theory for itinerant quantum
helimagnets. We will do so by deriving a quantum Ginzburg-Landau theory whose
coefficients are given in terms of microscopic electronic correlation
functions. Such a theory has two advantages over a purely phenomenological
treatment based on symmetry arguments alone. First, it allows for a
semi-quantitative analysis, since the coefficients of the Ginzburg-Landau
theory can be expressed in terms of microscopic parameters. Second, it derives
all of the ingredients necessary for calculating the thermodynamic and
transport properties of an itinerant helimagnet in the ordered phase using
many-body perturbation theory techniques.\cite{paper_II}

The organization of the remainder of this paper is as follows. In Sec.\
\ref{sec:II} we use an analogy with chiral liquid crystals to make an educated
guess about the wave vector dependence of fluctuations in helimagnets, and
employ time-dependent Ginzburg-Landau theory to find the dynamics. In Sec.\
\ref{sec:III} we derive the static properties from a classical Ginzburg-Landau
theory. In Sec. \ref{sec:IV} we start with a microscopic quantum mechanical
description and derive an effective quantum theory for chiral magnets. We then
show that all of the qualitative results obtained from the simple arguments in
Sec.\ \ref{sec:II} follow from this theory, with the additional benefit that
parameter values can be determined semi-quantitatively. We conclude in Sec.\
\ref{sec:V} with a summary and a discussion of our results. Some technical
details are relegated to three appendices.

\section{Simple physical arguments, and results}
\label{sec:II}

Helimagnets are not the only macroscopic systems that display chirality,
another examples are cholesteric liquid crystals whose director order parameter
is arranged in a helical pattern analogous to that followed by the
magnetization in a helimagnet.\cite{Chaikin_Lubensky_1995_ch_6.2} There are
some important differences between magnets and liquid crystals. For instance,
the two orientations of the director order parameter in the latter are
equivalent, which necessitates a description in terms of a rank-two tensor,
rather than a vector as in magnets.\cite{DeGennes_Prost_1993} Also, the
chirality in cholesteric liquid crystals is a consequence of the chiral
properties of the constituting molecules, whereas in magnets it is a result of
interactions between the electrons and the atoms of the underlying lattice.
However, these differences are not expected to be relevant for some basic
properties of the Goldstone mode that must be present in the helical state of
either system.\cite{dynamics_footnote} We will therefore start by using the
known hydrodynamic properties of cholesteric liquid crystals to motivate a
guess of the nature of the Goldstone mode in helimagnets. In Sec.\
\ref{sec:III} we will see that the results obtained in this way are indeed
confirmed by an explicit calculation. The arguments employed in this section
are phenomenological in nature and very general. We therefore expect them to
apply equally to classical helimagnets and to quantum helimagnets at $T=0$, as
is the case for analogous arguments for ferromagnetic and antiferromagnetic
magnons.

\subsection{Statics}
\label{subsec:II.A}

Consider a classical magnet with an order parameter field ${\bm M}({\bm x})$
and an action\cite{Bak_Jensen_1980, Nakanishi_et_al_1980}
\bea
S[{\bm M}] &=& \int d{\bm x}\,\left[\frac{r}{2}\,{\bm M}^2({\bm x}) +
\frac{a}{2}\,\left(\nabla{\bm M}({\bm x})\right)^2 \right.
\nonumber\\
&& \left. \hskip -40pt +\ \frac{c}{2}\,{\bm M}({\bm
x})\cdot\left({\bm\nabla}\times{\bm M}({\bm x})\right) +
\frac{u}{4}\,\left({\bm M}^2({\bm x})\right)^2\right].
\label{eq:2.1}
\eea
This is a classical $\phi^4$-theory with a chiral term with coupling constant
$c$. Physically, $c$ is proportional to the spin-orbit coupling strength
$g_{\text{\,SO}}$. The expectation value of ${\bm M}$ is proportional to the
magnetization, and it is easy to see that a helical field configuration
constitutes a saddle-point solution of the action given by Eq.\ (\ref{eq:2.1}),
\bse
\label{eqs:2.2}
\bea
{\bm M}_{\text{sp}}({\bm x}) &=& m_0\,\left({\bm e}_1\,\cos {\bm q}\cdot{\bm x}
+ {\bm e}_2\,\sin {\bm q}\cdot{\bm x}\right)
\label{eq:2.2a}\\
 &=& m_0\,\left(\cos qz, \sin qz, 0\right).
\label{eq:2.2b}
\eea
\ese
In Eq.\ (\ref{eq:2.2a}), ${\bm e}_1$ and ${\bm e}_2$ are two unit vectors that
are perpendicular to each other and to the pitch vector ${\bm q}$. The
chirality of the dreibein $\{{\bm q},{\bm e}_1,{\bm e}_2\}$ reflects the
chirality of the underlying lattice structure and is encoded in the coefficient
$c$ in Eq.\ (\ref{eq:2.1}), with the sign of $c$ determining the handedness of
the chiral structure. In Eq.\ (\ref{eq:2.2b}) we have chosen a coordinate
system such that $\{{\bm e}_1, {\bm e}_2, {\bm q}/q\} = \{{\hat x},{\hat
y},{\hat z}\}$, a choice we will use for all explicit calculations. We will
further choose, without loss of generality, $c>0$. The free energy is minimized
by $q = c/2a$, and the pitch wave number is thus proportional to
$g_{\text{\,SO}}$.

Now consider fluctuations about this saddle point. An obvious guess for the
soft mode associated with the ordered helical state are phase fluctuations of
the form
\bea
{\bm M}({\bm x}) &=& m_0\,\left(\cos (qz + \phi({\bm x})), \sin (qz + \phi({\bm
x})), 0\right)
\nonumber\\
&&\hskip -30pt = {\bm M}_{\text{sp}}({\bm x}) + m_0\,\phi({\bm x})\,\left(-\sin
qz , \cos qz , 0\right) + O(\phi^2).
\nonumber\\
\label{eq:2.3}
\eea
These phase fluctuations are indeed soft; by substituting Eq.\ (\ref{eq:2.3})
in Eq.\ (\ref{eq:2.1}) one finds an effective action $S_{\text{eff}}[\phi] =
\text{const.}\int d{\bm x}\,\left({\bm\nabla}\phi({\bm x})\right)^2$. However,
this cannot be the correct answer, which can be seen as
follows.\cite{Chaikin_Lubensky_1995_ch_6.3} Consider a simple rotation of the
planes containing the spins such that their normal changes from $(0,0,q)$ to
$(\alpha_1,\alpha_2,q)$, which corresponds to a phase fluctuation $\phi({\bm
x}) = \alpha_1\,x + \alpha_2\,y$. This cannot cost any energy, yet
$({\bm\nabla}\phi)^2 = \alpha_1^2 +\, \alpha_2^2 \neq 0$ for this particular
phase fluctuation. The problem is the dependence of the effective action on
${\bm\nabla}_{\perp}\phi$, where ${\bm\nabla} =
({\bm\nabla}_{\perp},\partial_z)$. The soft mode must therefore be some
generalized phase $u({\bm x})$ with a schematic structure
\be
u({\bm x}) \sim \phi({\bm x}) + \nabla_{\perp}\varphi({\bm x}),
\label{eq:2.4}
\ee
where $\varphi({\bm x})$ represents the $z$-component of the order parameter
vector ${\bm M}({\bm x})$. The lowest order dependence on perpendicular
gradients allowed by rotational invariance is ${\bm\nabla}_{\perp}^2 u$, and
the extra term in $u$ proportional to $\nabla_{\perp}$ will ensure that this
requirement is fulfilled. The correct effective action thus is expected to have
the form
\be S_{\text{eff}}[u] = \frac{1}{2}\int d{\bm x}\,\left[c_z
   \left(\partial_z u({\bm x})\right)^2
+ c_{\perp}\left({\bm\nabla}_{\perp}^2 u({\bm x})\right)^2/q^2\right],
\label{eq:2.5}
\ee
where $c_z$ and $c_{\perp}$ are elastic constants. The Goldstone mode
corresponding to helical order must therefore have an anisotropic dispersion
relation: it will be softer in the direction perpendicular to the pitch vector
than in the longitudinal direction.\cite{fluctuations_footnote} Separating wave
vectors ${\bm k} = ({\bm k}_{\perp},k_z)$ into transverse and longitudinal
components, the longitudinal wave number will scale as the transverse wave
number squared, $k_z \sim {\bm k}_{\perp}^2/q$. The factor $1/q^2$ in the
transverse term in Eq.\ (\ref{eq:2.5}), which serves to ensure that the
constants $c_z$ and $c_{\perp}$ have the same dimension, is the natural length
scale to enter at this point, since a nonzero pitch wave number is what is
causing the anisotropy in the first place. A detailed calculation for
cholesteric liquid crystals \cite{Lubensky_1972} shows that this is indeed the
correct answer, and we will see in Sec.\ \ref{sec:III} that the same is true
for helimagnets.

\subsection{Dynamics}
\label{subsec:II.B}

In order to determine the dynamics of the soft mode in a simple
phenomenological fashion we utilize the framework of time-dependent
Ginzburg-Landau theory.\cite{Ma_1976_ch_XIII} Within this formalism, the
kinetic equation for the time-dependent generalization of the magnetization
field ${\bm M}$ reads
\bea
\partial_t {\bm M}({\bm x},t) &=& -\gamma\,{\bm M}({\bm x},t)\times\frac{\delta
S}{\delta {\bm M}({\bm x})}\biggr\vert_{{\bm M}({\bm x},t)}\hskip 40pt
\nonumber\\
&& \hskip -40pt - \int d{\bm y}\,D({\bm x}-{\bm y})\,\frac{\delta S}{\delta
{\bm M}({\bm y})}\biggr\vert_{{\bm M}({\bm y},t)} +\, {\bm\zeta}({\bm x},t),
\label{eq:2.6}
\eea
with $\gamma$ a constant. The first term describes the precession of a magnetic
moment in the field provided by all other magnetic moments, $D$ is a
differential operator describing dissipation that we will specify in Sec.\
\ref{subsec:II.C}, and ${\bm\zeta}$ is a random Langevin force with zero mean,
$\langle{\bm\zeta}({\bm x},t)\rangle = 0$, and a second moment consistent with
the fluctuation-dissipation theorem.

Now assume an equilibrium state given by Eq.\ (\ref{eq:2.2b}). In considering
deviations from the equilibrium state we must take into account both the
generalized phase modes at wave vector ${\bm q}$, which are soft since they are
Goldstone modes, and the modes at zero wave vector, which are soft due to spin
conservation. The latter we denote by ${\bm m}({\bm x},t)$, and for the former
we use Eq.\ (\ref{eq:2.3}),\cite{phase_footnote}
\bea
{\bm M}({\bm x},t) &=& {\bm M}_{\text{sp}}({\bm x}) + {\bm m}({\bm x},t)
\nonumber\\
&& +\, m_0\,u({\bm x},t)(-\sin qz,\cos qz,0).
\label{eq:2.7}
\eea
The action for $u$ is the effective action given by Eq.\ (\ref{eq:2.5}), and
the action for ${\bm m}$ is a renormalized Ginzburg-Landau action of which we
will need only the Gaussian mass term. We thus write
\be
S[{\bm m},u] = \frac{r_0}{2}\int d{\bm x}\ {\bm m}^2({\bm x}) +
S_{\text{eff}}[u].
\label{eq:2.8}
\ee
The mass $r_0$ of the zero-wave number mode that appears here and in the
remainder of this section is different from the coefficient $r$ in Eq.\
(\ref{eq:2.1}), and we assume $r_0>0$.\cite{mass_footnote}

\begin{figure}[t,h]
\vskip -0mm
\includegraphics[width=8.5cm]{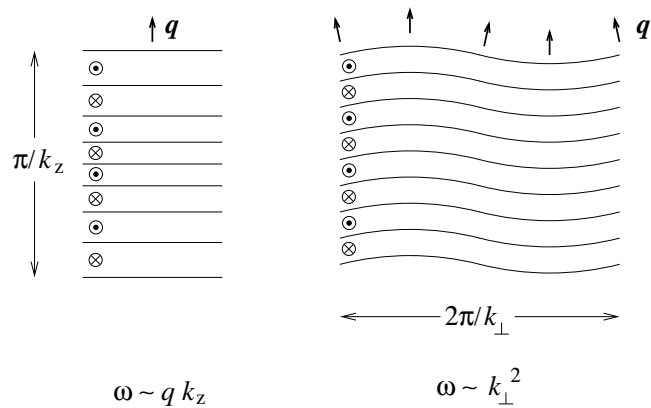}
\caption{Sketch of a longitudinal (${\bm k}\vert\vert{\bm q}$, left panel) and
transverse (${\bm k}\perp{\bm q}$, right panel) helimagnon. The solid lines
delineate planes of spins pointing out of (dotted circle) or into (crossed
circle) the paper plane. }
\label{fig:1}
\end{figure}

We now use the kinetic equation (\ref{eq:2.6}) to calculate the average
deviations $\langle{\bm m}({\bm x},t)\rangle$ and $\langle u({\bm x},t)\rangle$
from the equilibrium state. For simplicity we suppress both the averaging
brackets and the explicit time dependence in our notation, and for the time
being we neglect the dissipative term. With summation over repeated indices
implied, Eq.\ (\ref{eq:2.6}) yields
\bse
\label{eqs:2.9}
\bea
\partial_t\,m_3({\bm x}) &=& -\gamma\,\epsilon_{3ij}\, M_{\text{sp}}^i({\bm x}) \int
d{\bm y}\ \frac{\delta S}{\delta u({\bm y})}\,\frac{\delta u({\bm y})}{\delta
M_j({\bm x})} \nonumber\\
&=& -\gamma\,m_0\int d{\bm y}\ \frac{\delta S}{\delta u({\bm y})}\,\biggl[\cos
qz\,\frac{\partial u({\bm y})}{\partial M_y({\bm x})} \nonumber\\
&&\hskip 60pt  -\sin qz\,\frac{\partial u({\bm y})}{\partial M_x({\bm
x})}\biggr]\,.
\label{eq:2.9a}
\eea
and by using Eq.\ (\ref{eq:2.7}) in the identity
\begin{equation*}
\delta({\bm x}-{\bm y}) = \int d{\bm z}\ \frac{\delta u({\bm x})}{\delta
M_i({\bm z})}\,\frac{\delta M_i({\bm z})}{\delta u({\bm y})}
\end{equation*}
we
find
\be
m_0\left(-\sin qz\,\frac{\delta u({\bm x})}{\delta M_x({\bm y})} + \cos
qz\,\frac{\delta u({\bm x})}{\delta M_y({\bm y})}\right) = \delta({\bm x}-{\bm
y}).
\label{eq:2.9b}
\ee
\ese
Using Eq.\ (\ref{eq:2.9b}) in Eq.\ (\ref{eq:2.9a}) eliminates the integration,
and using Eqs. (\ref{eq:2.8}) and (\ref{eq:2.5}) we find a relation between
$m_3$ and $u$,
\be
\partial_t m_3({\bm x}) = -\gamma\,\frac{\delta S_{\text{eff}}}{\delta u({\bm x})} =
-\gamma\left(-c_z\partial_z^2 +
c_{\perp}{\bm\nabla}_{\perp}^4/q^2\right)\,u({\bm x}).
\label{eq:2.10}
\ee
A second relation is obtained from the identity
\begin{equation*}
\partial_t M_1({\bm x}) = \int d{\bm y}\,\frac{\delta M_1({\bm
x})}{\delta u({\bm y})}\,\partial_t u({\bm y}).
\end{equation*}
By applying Eq.\ (\ref{eq:2.6}) to the left-hand side, and using Eqs.\
(\ref{eq:2.8}) and (\ref{eq:2.7}) we obtain
\begin{equation*}
\int d{\bm y}\,\frac{\delta M_1({\bm x})}{\delta u({\bm y})}\,\partial_t u({\bm
y}) = \gamma\,r_0 \int d{\bm y}\,\frac{\delta M_1({\bm x})}{\delta u({\bm
y})}\,m_3({\bm y})
\end{equation*}
or
\be
\partial_t u({\bm x}) = \gamma\,r_0\,m_3({\bm x}).
\label{eq:2.11}
\ee
Combining Eqs.\ (\ref{eq:2.10}, \ref{eq:2.11}) we find a wave equation
\be
\partial_t^2 u({\bm x}) = -\gamma^2\,r_0\,\left(-c_z\partial_z^2 +
c_{\perp}{\bm\nabla}_{\perp}^4/q^2\right)\,u({\bm x}).
\label{eq:2.12}
\ee
This is the equation of motion for a harmonic oscillator with a resonance
frequency
\be
\omega_0({\bm k}) = \gamma\,r_0^{1/2}\sqrt{c_z\,k_z^2 + c_{\perp}\,{\bm
k}_{\perp}^4/q^2}
\label{eq:2.13}
\ee
and a susceptibility
\be
\chi_0 = \frac{1}{\omega_0^2({\bm k}) - \omega^2}\ .
\label{eq:2.14}
\ee
We thus have a propagating mode, the helimagnon, with an anisotropic dispersion
relation: for wave vectors parallel to the pitch vector ${\bm q}$ the
dispersion is linear, as in an antiferromagnet, while for wave vectors
perpendicular to ${\bm q}$ it is quadratic. Fluctuations transverse with
respect to the pitch vector are thus softer than longitudinal ones. The nature
of the excitation corresponding to the longitudinal and transverse helimagnon,
respectively, is shown in Fig.\ \ref{fig:1}.

For later reference we note that for determining the static properties of the
helimagnon it sufficed to discuss the phase modes at wave vector ${\bm q}$,
while the dynamics are generated by a coupling between the phase modes and the
modes at zero wave vector. This observation gives an important clue for the
correct structure of the microscopic theory we will develop in Sec.\
\ref{sec:IV}.

\subsection{Damping}
\label{subsec:II.C}

In order to investigate the damping of the mode we need to take into account
the dissipative term in the master equation (\ref{eq:2.6}) which we have
neglected so far. Usually, in the case of a conserved order parameter, the
damping operator $D$ in Eq.\ (\ref{eq:2.6}) is proportional to a gradient
squared.\cite{Ma_1976_ch_XIII} In the present case, however, one expects an
anisotropic differential operator, with different prefactors for the
longitudinal and transverse parts, respectively. We will see in Sec.\
\ref{sec:IV} that in the particular model we will consider the transverse part
of the gradient squared has a zero prefactor. We thus write
\be
D({\bm x}-{\bm y}) = \Gamma\,\delta({\bm x}-{\bm y})\,\partial_z^2,
\label{eq:2.15}
\ee
where $\Gamma$ is a damping coefficient. Going through the derivation in the
previous subsection again, we see that Eq.\ (\ref{eq:2.11}) remains unchanged
except for gradient corrections to the right-hand side. Equation
(\ref{eq:2.10}), on the other hand, acquires an additional term that is of the
same order as the existing ones,
\bea
\partial_t m_3({\bm x}) &=&
-\gamma\left(-c_z\partial_z^2 +
c_{\perp}{\bm\nabla}_{\perp}^4/q^2\right)\,u({\bm x})
\nonumber\\
&& - r_0\Gamma\partial_z^2 m_3({\bm x}).
\label{eq:2.16}
\eea
Together with Eq.\ (\ref{eq:2.11}) this leads to an equation of motion for $u$
given by
\bea
\partial_t^2 u({\bm x}) &=& -\gamma^2\,r_0\,\left(-c_z\partial_z^2 +
c_{\perp}{\bm\nabla}_{\perp}^4/q^2\right)\,u({\bm x})
\nonumber\\
&& - r_0\Gamma\partial_z^2\partial_t\,u({\bm x}). \label{eq:2.17}
\eea
This corresponds to a damped harmonic oscillator with a susceptibility
\bse
\label{eqs:2.18}
\be
\chi = \frac{1}{\omega_0^2({\bm k}) - \omega^2  - i\omega\,\gamma({\bm k})}\ ,
\label{eq:2.18a}
\ee
where the damping coefficient is given by
\be
\gamma({\bm k}) = r_0\Gamma k_z^2.
\label{eq:2.18b}
\ee
\ese

Now recall that we are interested in systems where the magnetization is caused
by itinerant electrons. In a system without any elastic scattering due to
impurities, and at zero temperature, the coefficient $\Gamma$, which physically
is related to a generalized viscosity of the electron fluid, is itself wave
number dependent and diverges for ${\bm k}\to 0$ as $\Gamma \propto 1/\vert{\bm
k}\vert$. This leads to a damping coefficient
\bse
\label{eqs:2.19}
\be
\gamma({\bm k}\to 0) \propto k_z^2/\vert{\bm k}\vert.
\label{eq:2.19a}
\ee
We see that $\gamma({\bm k})$ scales as $k_z$ (for $k_z^2 \gg {\bm
k}_{\perp}^4/q^2$), or as $k_z^{3/2}$ (for $k_z^2 \ll {\bm k}_{\perp}^4/q^2$),
while the resonance frequency $\omega_0$ scales as $k_z$. If the prefactor of
the damping coefficient is not too large, we thus have $\gamma({\bm k}) <
\omega_0({\bm k})$ for all ${\bm k}$ and the mode is propagating. Any amount of
quenched disorder will lead to $\Gamma$ being finite at zero wave vector, and
hence to
\be
\gamma({\bm k}\to 0) \propto k_z^2.
\label{eq:2.19b}
\ee
\ese
In this case, the mode is propagating at all wave vectors irrespective of the
prefactor.

\subsection{Physical Spin Susceptibility}
\label{subsec:II.D}

The physical spin susceptibility $\chi_{\text{s}}$, which is directly
measurable, is given in terms of the order-parameter correlation function. The
transverse (with respect to ${\bm q})$ components of $\chi_{\text{s}}$ are
given by the correlations of the phase $\phi$ in Eq.\ (\ref{eq:2.3}), and are
thus directly proportional to the Goldstone mode. In a schematic notation,
which ignores the fact that $\phi$ at zero wave vector corresponds to a
magnetization fluctuation at wave vector $\pm{\bm q}$, we thus expect
\bse
\label{eqs:2.20}
\be
\chi_{\text{s}}^{\perp}({\bm k},\omega) \propto \frac{1}{\omega_0^2({\bm k}) -
\omega^2 - i\omega\,\gamma({\bm k})}\ .
\label{eq:2.20a}
\ee
The longitudinal component will, by Eq.\ (\ref{eq:2.4}), carry an additional
factor of ${\bm k}_{\perp}^2$, and is thus expected to have the structure
\be
\chi_{\text{s}}^{\text{L}}({\bm k},\omega) \propto \frac{{\bm
k}_{\perp}^2}{\omega_0^2({\bm k}) - \omega^2 - i\omega\,\gamma({\bm k})}\ .
\label{eq:2.20b}
\ee
\ese
Since $\omega \sim k_z \sim {\bm k}_{\perp}^2$ in a scaling sense, we see that
the transverse susceptibility, $\chi_{\text{s}}^{\perp} \sim 1/\omega^2$, is
softer than the longitudinal one, $\chi_{\text{s}}^{\text{L}} \sim 1/\omega$.

\subsection{Effects of broken rotational and translational invariance}
\label{subsec:II.E}

For the arguments given so far, the rotational symmetry of the action $S[{\bm
M}]$, Eq.\ (\ref{eq:2.1}), i.e., the invariance under simultaneous rotations in
real space and spin space, played a crucial role. Since the underlying lattice
structure of a real magnet breaks this symmetry, it is worthwhile to consider
the consequences of this effect.

In a system with a cubic lattice like MnSi, the simplest term that breaks the
rotational invariance is of the form\cite{Bak_Jensen_1980,
Nakanishi_et_al_1980}
\bea
S_{\text{cubic}}[{\bm M}] &=& \frac{a_1}{2}\int d{\bm x}\
\left[\left(\partial_x M_x({\bm x})\right)^2 + \left(\partial_y M_y({\bm
x})\right)^2\right.
\nonumber\\
&& \hskip 50pt + \left. \left(\partial_z M_z({\bm x})\right)^2\right].
\label{eq:2.21}
\eea
Other anisotropic terms with a cubic symmetry (see Appendix \ref{app:X} for a
complete list) have qualitatively the same effect. In Eq.\ (\ref{eq:2.21}),
$a_1\propto g_{\text{\,SO}}^2$, with $g_{\text{\,SO}}$ the spin-orbit coupling
strength (see Sec.\ \ref{subsec:II.A}). On dimensional grounds, we thus have
$a_1 = b\, q^2 a$, with $b$ a number, and $a$ the coefficient of the gradient
squared term in Eq.\ (\ref{eq:2.1}). $b\neq 0$ leads to a pinning of the helix
pitch vector in (1,1,1) or equivalent directions (for $b<0$), or in (1,0,0) or
equivalent directions (for $b>0$).\cite{Bak_Jensen_1980, Nakanishi_et_al_1980}
In addition, it invalidates the argument in Sec.\ \ref{subsec:II.A} that there
cannot be a $({\bm\nabla}_{\perp}\phi)^2$ term in the effective action.
However, the action is still translationally invariant, so a constant phase
shift cannot cost any energy. To Eq.\ (\ref{eq:2.1}) we thus need to add a term
\be
S_{\text{eff}}^{\text{cubic}}[u] = \frac{b\,q^2 a}{2}\int d{\bm x}\
\left({\bm\nabla}_{\perp} u({\bm x})\right)^2.
\label{eq:2.22}
\ee
This changes the soft-mode frequency, Eq.\ (\ref{eq:2.13}), to
\be
\omega_0({\bm k}) = \gamma r_0^{1/2}\sqrt{c_z\,k_z^2 + b\,a\,q^2{\bm
k}_{\perp}^2 + c_{\perp}{\bm k}_{\perp}^4/q^2}.
\label{eq:2.23}
\ee
We note that, due to the weakness of the spin-orbit coupling, $a\,q^2\ll 1$,
and therefore the breaking of the rotational symmetry is a very small effect.
In MnSi, where the pitch wave length is on the order of 200\AA, while $a^{1/2}$
is on the order of a few \AA\ at most, the presence of the ${\bm k}_{\perp}^2$
term is not observable with the current resolution of neutron scattering
experiments, and we will ignore this term in the remainder of this paper.

The above considerations make it clear that the Goldstone mode is due to the
spontaneous breaking of translational invariance, rather than rotational
invariance in spin space. Consistent with this, there is only one Goldstone
mode, as the helical state is still invariant under a two-parameter subgroup of
the original three-parameter translational
group.\cite{number_soft_modes_footnote} In this sense, the helimagnon is more
akin to phonons than to ferromagnetic magnons. Let us briefly discuss the
effect of the ionic lattice on this symmetry, as the helix can be pinned by the
periodic lattice potential and therefore one expects a gap in the magnetic
excitation spectrum. To estimate the size of the gap, we investigate the
low-energy theory taking into account only slowly varying modes with $|\bm
k-\bm q| \lesssim q$. In a periodic lattice, momentum is conserved up to
reciprocal lattice vectors $\bm G_j$. The leading term which breaks
translational invariance, ${\bm M}_{\bm k} \to \bm M_{\bm k} e^{i \bm k \bm
r_0}$ therefore is of the form
\be
S_n=\sum_{\bm k_1,\ldots,\bm k_n,j } V_{\{\bm k_l\},\bm G_j}M_{\bm k_1} M_{\bm
k_2} \ldots M_{\bm k_n} \, \delta\!\left(\sum_i \bm k_i - \bm G_j\right)
\label{eq:2.24}
\ee
where $V$ parameterizes the momentum dependent coupling strength (and we have
omitted vector indices). Within the low-energy theory, all momenta are of order
of $q$. Therefore, umklapp scattering can  only take place for $n \gtrsim G/q$.
In the case of MnSi, where $G/q \approx 40$, one therefore needs a process
proportional to $M^{40}$ to create a finite gap! It is difficult to estimate
the precise size of the gap which depends crucially, for example, on the
commensuration of the helix with the underlying lattice. However, the resulting
gap will in any case be unobservably small as it is exponentially suppressed by
the large parameter $G/q \propto 1/g_{\text{\,SO}}$.

\section{Nature of the Goldstone mode in classical chiral magnets}
\label{sec:III}

One of our goals is to derive from a microscopic theory the results one expects
based on the simple considerations in Sec.\ \ref{sec:II}. As a first step, we
show that the phenomenological action for classical helimagnets given by Eq.\
(\ref{eq:2.1}) does indeed result in the effective elastic theory given by Eq.\
(\ref{eq:2.5}). A derivation from a microscopic quantum mechanical Hamiltonian
will be given in Sec.\ \ref{sec:IV}.

The classical $\phi^4$-theory with a chiral term represented by the action
given in Eq.\ (\ref{eq:2.1}) can be analyzed in analogy to the action for
chiral liquid crystals.\cite{Lubensky_1972} In the magnetic case the chiral
term with coupling constant $c$ is of the form first proposed by Dzyalishinkski
\cite{Dzyaloshinski_1958} and Moriya,\cite{Moriya_1960} who showed that it is a
consequence of the spin-orbit interaction in crystals that lack spatial
inversion symmetry.

\subsection{Saddle-point solution}
\label{subsec:III.A}

The saddle-point equation, $\delta S/\delta M_i({\bm x}) = 0$, reads
\bse
\label{eqs:3.1}
\be
\left(r - a{\bm\nabla}^2 + c{\bm\nabla}\!\times +\, u\,{\bm M}^2({\bm
x})\right)\,{\bm M}({\bm x}) = 0,
\label{eq:3.1a}
\ee
and the free energy density in saddle-point approximation is given by
\be
f_{\text sp} = \frac{T}{V}\,S[{\bm M}_{\text sp}],
\label{eq:3.1b}
\ee
\ese
with ${\bm M}_{\text sp}$ a solution of Eq.\ (\ref{eq:3.1a}).

The helical field configuration given by Eqs.\ (\ref{eqs:2.2}) with an
amplitude $m_0^2 = -(r + aq^2 - cq)/u$ is a solution for any value of $q$. The
physical value of $q$ is determined from the requirement that the free energy
must be minimized, which yields
\bse
\label{eqs:3.2}
\be
q = c/2a.
\label{eq:3.2a}
\ee
and
\be
m_0 = \frac{1}{\sqrt{u}}\,\left(c^2/4a - r\right)^{1/2}.
\label{eq:3.2b}
\ee
\ese
The zero solution, $m_0=0$, is unstable with respect to the helical solution
for all $r < c^2/4a$, and for $c\neq 0$ the ferromagnetic solution $q=0$,
$m_0^2 = -r/u$ is always unstable with respect to the helical one since one can
always gain energy by making $q\neq 0$ due to the linear momentum dependence of
the chiral term ${\bm M}\cdot({\bm\nabla}\times{\bm M})$.

\subsection{Gaussian fluctuations}
\label{subsec:III.B}

\subsubsection{Disordered phase}
\label{subsubsec:III.B.1}

In the disordered phase, $r > c^2/4a$, the Gaussian propagator is easily found
by inverting the quadratic form in Eq. (\ref{eq:2.1}),
\bea
\langle M_i({\bm k})\,M_j({\bm p})\rangle &=& \delta_{{\bm k},-{\bm
p}}\,\frac{1}{(r + a q^2)^2 - c^2 q^2}
\nonumber\\
&& \hskip -80pt \times \left[\delta_{ij} (r + a q^2) +
\epsilon_{ij\,l}\,ic\,k_{\,l} - k_ik_j\,\frac{c^2}{r + a q^2}\right].
\label{eq:3.3}
\eea
The structure of the prefactor in this expression is consistent with the
conclusion of Sec.\ \ref{subsec:III.A}: For $r > c^2/4a$, the denominator $N(q)
= (r + a q^2)^2 - c^2 q^2$ is minimized by $q=0$, and $N(q)$ has no zeros in
this regime. $N(q)$ first reaches zero at $r = c^2/4a$ and $q=c/2a$, and the
disordered phase is unstable for all $r < c^2/4a$. The quantum-critical
fluctuations in the disordered phase have been discussed by Schmalian and
Turlakov.\cite{Schmalian_Turlakov_2004}

\subsubsection{Ordered phase}
\label{subsubsec:III.B.2}

In the ordered phase the determination of the Gaussian fluctuations is more
complicated. Let us parameterize the order parameter field as follows,
\bse
\label{eqs:3.4}
\be
{\bm M}({\bm x}) = \left(m_0 + \delta m({\bm x})\right)
                             \left(\begin{array}{c} \cos(qz + \phi({\bm x})) \\
                                                    \sin(qz + \phi({\bm x})) \\
                                                    \varphi({\bm x})
                                   \end{array}\right),
\label{eq:3.4a}
\ee
where $\phi({\bm x})$, $\varphi({\bm x})$, and $\delta m({\bm x})$ describe
small fluctuations about the saddle-point solution. Fluctuations of the norm of
${\bm M}$ one expects to be massive, as they are in the ferromagnetic case, and
an explicit calculation confirms this expectation. We thus can keep the norm of
${\bm M}$ fixed, which means that $\delta m$ is quadratic in the small
fluctuation $\varphi$ and does not contribute to the Gaussian action. In order
to treat the $\phi$ and $\varphi$ fluctuations, it is useful to acknowledge
that, upon performing a Fourier transform, $\phi({\bm k}=0)$ corresponds to
taking ${\bm M}$ at ${\bm k}={\bm q}$, while $\varphi$ and ${\bm M}$ come at
the same wave number. We therefore write
\be
\varphi({\bm x}) = \varphi_1({\bm x})\,\sin qz + \varphi_2({\bm x})\,\cos qz,
\label{eq:3.4b}
\ee
\ese
where $\varphi_1$ and $\varphi_2$ are restricted to containing Fourier
components with $\vert{\bm k}\vert \ll q$.\cite{varphi_footnote} The Goldstone
mode is now expected to be a linear combination of $\phi$, $\varphi_1$, and
$\varphi_2$ at zero wave vector. If we expand Eq.\ (\ref{eq:3.4a}) to linear
order in $\phi \equiv \varphi_0$, substitute this in Eq.\ (\ref{eq:2.1}),
neglect rapidly fluctuating Fourier components proportional to $e^{inqz}$ with
$n\ge 2$, and use the equation of motion (\ref{eq:3.1a}), we obtain a Gaussian
action
\bse
\label{eqs:3.5}
\be
S^{(2)}[\varphi_i] = \frac{a\,m_0^2}{2}\sum_{\bm p}\sum_{i=0,1,2}
\varphi_i({\bm p})\,\gamma_{ij}({\bm p})\,\varphi_j(-{\bm p})
\label{eq:3.5a}
\ee
with a matrix
\be
\gamma({\bm p}) = \left(\begin{array}{ccc}
                              {\bm p}^{\,2} & -iqp_{\,y}             & -iqp_{\,x} \\
                                  iqp_{\,y} & q^2 + {\bm p}^{\,2}/2 & iqp_{\,z} \\
                                  iqp_{\,x} & -iqp_{\,z}            & q^2 + {\bm p}^{\,2}/2
                          \end{array}\right).
\label{eq:3.5b}
\ee
\ese
The corresponding eigenvalue equation reads
\bea
(\mu - {\bm p}^{\,2})(q^2 + {\bm p}^{\,2}/2 - \mu)^2 &+& q^2 {\bm
p}_{\perp}^{\,2} (q^2 + {\bm p}^{\,2}/2 - \mu)
\nonumber\\
&+& q^2 p_z^{\,2} ({\bm p}^{\,2} - \mu) = 0.
\label{eq:3.6}
\eea
We see that at ${\bm p}=0$ there is one eigenvalue $\mu_1 = 0$ and a doubly
degenerate eigenvalue $\mu_{2,3} = q^2$. As expected, there thus is one soft
(Goldstone) mode in the ordered phase. The behavior at nonzero wave vector is
easily determined by solving Eq. (\ref{eq:3.6}) perturbatively. The degeneracy
of $\mu_2$ and $\mu_3$ is lifted, $\mu_{2,3}({\bm p}\to 0) = q^2 \pm qp_z$, and
for $\mu_1$ we find
\bse
\label{eqs:3.7}
\be
\mu_1({\bm p}\to 0) = p_z^{\,2} + {\bm p}_{\perp}^{\,4}/2q^2 + O(p_z^{\,2}{\bm
p}_{\perp}^{\,2}).
\label{eq:3.7a}
\ee
The corresponding eigenvector is
\bea
v_1({\bm p}) = \phi({\bm p}) &-& i\,(p_y/q)[1 + O({\bm
p}_{\perp}^{\,2})]\,\varphi_1({\bm p})
\nonumber\\
&-& i\,(p_x/q)[1 + O({\bm p}_{\perp}^{\,2})]\,\varphi_2({\bm p}).
\label{eq:3.7b}
\eea
It has the property
\be
\langle v_1({\bm p})\,v_1(-{\bm p})\rangle = 1/a\,m_0^2\,\mu_1({\bm p}).
\label{eq:3.7c}
\ee
\ese

A comparison with Eq.\ (\ref{eq:2.5}) shows that the effective soft-mode action
has indeed the form that was expected from the analogy with chiral liquid
crystals. If we identify $\sqrt{a\,m_0^2}\,v_1({\bm x})$ with the generalized
phase $u({\bm x})$, the coupling constants in Eq.\ (\ref{eq:2.5}) are $c_z = 1$
and $c_{\perp} = 1/2$. Repeating the calculation in the presence of a term that
breaks the rotational symmetry, e.g., Eq.\ (\ref{eq:2.21}), yields a result
consistent with Eq.\ (\ref{eq:2.22}) or (\ref{eq:2.23}), with $b = O(1)$.

\section{Nature of the Goldstone Mode in Quantum Chiral Magnets}
\label{sec:IV}

We now turn to the quantum case. Our objective is to develop an effective
theory for itinerant helimagnets that is analogous to Hertz's theory for
itinerant ferromagnets.\cite{Hertz_1976} That is, starting from a microscopic
fermionic action we derive a quantum mechanical generalization of the classical
Ginzburg-Landau theory studied in the preceding section. The coefficients of
this effective quantum theory will be given in terms of electronic correlation
functions, which allows for a semi-quantitative analysis of the results. In
addition, it provides the building blocks for a treatment of quantum
helimagnets by means of many-body perturbation theory, which will allow to go
beyond the treatment at a saddle-point/Gaussian level employed in the present
paper. We will show that this theory has a helical ground state given by Eqs.\
(\ref{eqs:2.2}), and consider fluctuations about this state to find the
Goldstone modes.

\subsection{Effective action for an itinerant quantum chiral magnet}
\label{subsec:IV.A}

Consider a partition function
\bse
\label{eqs:4.1}
\be
Z = \int D[{\bar\psi},\psi]\ e^{S[{\bar\psi},\psi]}
\label{eq:4.1a}
\ee
given by an electronic action of the form
\be
S[{\bar\psi},\psi] = {\tilde S}_0[{\bar\psi},\psi] +
S_{\text{int}}^{\,\text{t}}.
\label{eq:4.1b}
\ee
\ese
Here $S_{\text{int}}^{\,\text{t}}$ describes the spin-triplet interaction.
${\tilde S}_0[{\bar\psi},\psi]$, which we will explicitly specify later,
contains all other parts of the action, and the action is a functional of
fermionic (i.e., Grassmann-valued) fields $\psi$ and ${\bar\psi}$. The
spin-triplet interaction we take to have the form
\bse
\label{eqs:4.2}
\be
S_{\text{int}}^{\,\text{t}} = \frac{1}{2}\int d{\bm x}\,d{\bm y} \int_0^{1/T}
d\tau\ n_{\text{s}}^i({\bm x},\tau)\,A_{ij}({\bm x}-{\bm
y})\,n_{\text{s}}^j({\bm y},\tau).
\label{eq:4.2a}
\ee
Here and in what follows summation over repeated spin indices is implied. ${\bm
x}$ and ${\bm y}$ denote the position in real space, $\tau$ is the imaginary
time variable, and $n_{\text{s}}^i({\bm x},\tau)$ denote the components of the
electronic spin-density field ${\bm n}_{\text{s}}({\bm x},\tau)$. The
interaction amplitude $A$ is given by
\be
A_{ij}({\bm x}-{\bm y}) = \delta_{ij}\,\Gamma_{\text t}\,\delta({\bm x}-{\bm
y}) + \epsilon_{ij\,k}\,C_{\,k}({\bm x}-{\bm y}).
\label{eq:4.2b}
\ee
The first term, with a point-like amplitude $\Gamma_{\text{t}}$, is the usual
Hubbard interaction. The second term involves a cross-product ${\bm n}_{\text
s}({\bm x})\times {\bm n}_{\text{s}}({\bm y})$ and cannot exist in a
homogeneous electron system, which in particular is invariant under spatial
inversions. Dzyaloshinski \cite{Dzyaloshinski_1958} and Moriya
\cite{Moriya_1960} have shown that such a term arises from the spin-orbit
interaction in lattices that lack inversion symmetry. After coarse graining, it
will then also be present in an effective continuum theory valid at length
scales large compared to the lattice spacing. In such an effective theory the
vector ${\bm C}({\bm x}-{\bm y})$ is conveniently expanded in powers of
gradients. The lowest-order term in the gradient expansion is
\be
{\bm C}({\bm x}-{\bm y}) = c\,\Gamma_{\text{t}}\,\delta({\bm x}-{\bm
y})\,{\bm\nabla} + O(\nabla^2),
\label{eq:4.2c}
\ee
with $c$ a constant. The ferromagnetic case \cite{Hertz_1976} can be recovered
by putting $c=0$. We now perform a Hubbard-Stratonovich transformation to
decouple the spin-triplet interaction. To linear order in the gradients the
inverse of the matrix $A$ has the same form as $A$ itself, viz.,
\be
A^{-1}_{ij}({\bm x}-{\bm y}) =
\delta_{ij}\,\frac{1}{\Gamma_{\text{t}}}\,\delta({\bm x}-{\bm y}) -
\epsilon_{ijk}\,\frac{c}{\Gamma_{\text{t}}}\,\delta({\bm x}-{\bm
y})\,\partial_k + O(\nabla^2),
\label{eq:4.2d}
\ee
\ese
with $\partial_k = \partial/\partial x_k$ a spatial derivative. The
Hubbard-Stratonovich transformation thus produces all of the terms one gets in
the ferromagnetic case, and in addition a term
\be
-\frac{1}{2}\,c\,\Gamma_{\text{t}} \int d{\bm x}\ {\bm M}({\bm
x},\tau)\cdot\left({\bm\nabla}\times{\bm M}({\bm x},\tau)\right) + O(\nabla^2),
\label{eq:4.3}
\ee
%\eea
where ${\bm M}$ is the Hubbard-Stratonovich field whose expectation value is
proportional to the magnetization. The partition function can then be written
in the following form.
\bea
Z &=& \int D[{\bar\psi},{\psi}]\ e^{{\tilde S}_0[{\bar\psi},\psi]} \int D[{\bm
M}]\ e^{-(\Gamma_{\text{t}}/2)\int dx\,{\bm M}(x)\cdot{\bm M}(x)}\,
\nonumber\\
&&\times\, e^{-c(\Gamma_{\text{t}}/2)\int dx\,{\bm
     M}(x)\cdot\left({\bm\nabla}\times{\bm M}(x)\right)}\,
     e^{\Gamma_{\text{t}}\int dx\,{\bm M}(x)\cdot{\bm n}_{\text{s}}(x)}.
\nonumber\\
\label{eq:4.4}
\eea
Here we have adopted a four-vector notation, $x\equiv ({\bm x},\tau)$ and $\int
dx \equiv \int d{\bm x}\int_0^{1/T}d\tau$.

Now we consider the ordered phase and write
\be
{\bm M}(x) = {\bm M}_{\text{sp}}({\bm x}) + \delta{\bm M}(x),
\label{eq:4.5}
\ee
with ${\bm M}_{\text{sp}}$ given by Eq.\ (\ref{eq:2.2b}). The parameters $m_0$
and $q$ which characterize $M_{\text{sp}}$ will still have to be determined. By
substituting Eq.\ (\ref{eq:4.5}) in Eq.\ (\ref{eq:4.4}) and formally
integrating out the fermions we can write the partition function
\bse
\label{eqs:4.6}
\be
Z = \int D[\delta{\bm M}]\ e^{-{\cal A}[\delta{\bm M}]}
\label{eq:4.6a}
\ee
with ${\cal A}$ an effective action for the order-parameter fluctuations,
\bea
{\cal A}[\delta{\bm M}] &=& -\ln Z_0 + \frac{\Gamma_{\text{t}}}{2}\,\int dx\
{\bm M}(x)\cdot{\bm M}(x)
\nonumber\\
&& + \frac{c\Gamma_{\text{t}}}{2}\int dx\ {\bm
M}(x)\cdot\left({\bm\nabla}\times{\bm M}(x)\right)
\nonumber\\
&& -\ln \left\langle e^{\Gamma_{\text{t}} \int dx\ \delta{\bm M}(x)\cdot{\bm
n}_{\text{s}}(x)}\right\rangle_{S_0}.
\label{eq:4.6b}
\eea
\ese
Here
\bse
\label{eqs:4.7}
\be
S_0[{\bar\psi},\psi] = {\tilde S}_0[{\bar\psi},\psi] + \Gamma_{\text{t}} \int
dx\ {\bm M}_{\text{sp}}({\bm x})\cdot{\bm n}_{\text{s}}(x)
\label{eq:4.7a}
\ee
is a reference ensemble action for electrons described by ${\tilde S}_0$ in an
effective external magnetic field
\be
{\bm H}({\bm x}) = \Gamma_{\text{t}}\,{\bm M}_{\text{sp}}({\bm x}).
\label{eq:4.7b}
\ee
Only the Zeeman term due to the effective external field is included in the
reference ensemble. $Z_0$ is the partition function of the reference ensemble,
\be
Z_0 = \int D[{\bar\psi},\psi]\ e^{S_0[{\bar\psi},\psi]},
\label{eq:4.7c}
\ee
\ese
and $\langle \ldots \rangle_{S_0}$ denotes an average with respect to the
action $S_0$.

The effective action ${\cal A}$ can be expanded in a Landau expansion in powers
of $\delta{\bm M}$. To quadratic order this yields
\bse
\label{eqs:4.8}
\bea
{\cal A}[\delta{\bm M}] &=& \int dx\ \Gamma_i^{(1)}(x)\,\delta M_i(x)
\nonumber\\
&& +\ \frac{1}{2} \int dx\,dy\ \delta M_i(x)\,\Gamma_{ij}^{(2)}(x,y)\,\delta
M_j(y)
\nonumber\\
&& +\ O(\delta M^3).
\label{eq:4.8a}
\eea
with vertices
\be
\Gamma_i^{(1)}(x) = \Gamma_{\text{t}}(1 - c\,q)\,M_{\text{sp}}^i(x) -
\Gamma_{\text{t}}\,\left\langle n_{\text{s}}^i(x)\right\rangle_{S_0}
\label{eq:4.8b}
\ee
and
\bea
\Gamma_{ij}^{(2)}(x,y) &=& \delta_{ij}\,\delta(x-y)\,\Gamma_{\text{t}} -\
\epsilon_{ijk}\,\delta(x-y)\,\Gamma_{\text{t}}\,c\,\partial_k
\nonumber\\
&& - \chi_0^{ij}(x,y)\,\Gamma_{\text{t}}^2.
\label{eq:4.8c}
\eea
Here
\be
\chi_0^{ij}(x,y) = \left\langle
n_{\text{s}}^i(x)\,n_{\text{s}}^j(y)\right\rangle_{S_0}^{\text{c}}
\label{eq:4.8d}
\ee
\ese
is the spin susceptibility in the reference ensemble. The superscript c in Eq.\
(\ref{eq:4.8d}) indicates that only connected diagrams contribute to this
correlation function.

\subsection{Properties of the reference ensemble}
\label{subsec:IV.B}

In order for the formalism developed in the previous subsection to be useful we
need to determine the properties of the reference ensemble. As we will see in
Sec.\ \ref{subsec:IV.E}, and in a forthcoming paper,\cite{paper_II} the
reference ensemble is not only necessary for the present formal developments,
but also forms the basis for calculating all of the thermodynamic and transport
properties of helimagnets. This is because the reference ensemble, rather than
just being a useful artifact, has a precise physical interpretation: it
incorporates long-range helical order in a fermionic action at a mean-field
level.

We first need to specify the action ${\tilde S}_0$. For simplicity, we neglect
the spin-singlet interaction contained in ${\tilde S}_0$ and consider free
electrons with a Green function
\bse
\label{eqs:4.9}
\be
G_0({\bm k},i\omega_n) = 1/(i\omega_n - \xi_{\bm k})
\label{eq:4.9a}
\ee
Here $\omega_n = 2\pi T (n+1/2)$ is a fermionic Matsubara frequency, and
\be
\xi_{\bm k} = {\bm k}^2/2m_{\text{e}} - \epsilon_{\text{F}}
\label{eq:4.9b}
\ee
\ese
with $m_{\text{e}}$ the effective mass of the electrons and
$\epsilon_{\text{F}}$ the chemical potential or Fermi energy. (Here we neglect
spin-orbit interaction effects discussed in Ref.\
\onlinecite{Fischer_Rosch_2005} as well as quenched disorder.) For later
reference we also define the Fermi wave number $k_{\text{F}} =
\sqrt{2m_{\text{e}}\epsilon_{\text{F}}}$, the Fermi velocity $v_{\text{F}} =
k_{\text{F}}/m_{\text{e}}$, and the density of states per spin on the Fermi
surface $N_{\text{F}} = k_{\text{F}}m_{\text{e}}/2\pi^2$ in the ensemble
${\tilde S}_0$.

\subsubsection{Equation of state}
\label{par:IV.B.1}

The equation of state can be determined from the requirement
\cite{Ma_1976_ch_IX.7}
\bse
\label{eqs:4.10}
\be
\langle \delta {\bm M}({\bm x}) \rangle = 0,
\label{eq:4.10a}
\ee
where $\langle\ldots\rangle$ denotes an average with respect to the effective
action $\cal{A}$. To zero-loop order, this condition reads
\be
(1 - c\,q)\,{\bm M}_{\text{sp}}({\bm x}) - \langle {\bm
n}_{\text{s}}(x)\rangle_{S_0} = 0.
\label{eq:4.10b}
\ee
The zero-loop order or mean-field equation of state is thus determined by the
magnetization of the reference ensemble induced by the effective external field
$\Gamma_{\text{t}}\,{\bm M}_{\text{sp}}({\bm x})$, Eq.\ (\ref{eq:4.7b}). The
latter is given by the effective field times a generalized Lindhardt function.
The result is
\bea
1 - c\,q &=& - 2\,\Gamma_{\text{t}}\,\frac{1}{V}\sum_{\bm p}
\nonumber\\
&&\hskip -20pt \times T\sum_{i\omega_n} \frac{1}{G_0^{-1}({\bm
p},i\omega_n)\,G_0^{-1}({\bm p}-{\bm q},i\omega_n) - \lambda^2}\ ,
\nonumber\\
\label{eq:4.10c}
\eea
where $\lambda = m_0\,\Gamma_{\text{t}}$ is the exchange splitting or Stoner
gap. Notice that this provides only one relation between $\lambda$ and the
pitch wave number $q$. The latter still has to be determined from minimizing
the free energy, as in the classical case, Sec.\ \ref{subsec:III.A}.
\ese

\subsubsection{Green function}
\label{subsubsec:IV.B.2}

The basic building block for correlation functions of the reference ensemble is
the Green function associated with the action $S_0$, Eq.\ (\ref{eq:4.7a}). With
${\tilde S}_0$ as specified above, the latter reads explicitly
\be
S_0[{\bar\psi},\psi] = \int dx\,dy\ {\bar\psi}(x)\,G^{-1}(x,y)\,\psi( y).
\label{eq:4.11}
\ee
with an inverse Green function
\bse
\label{eqs:4.12}
\bea
G^{-1}(x,y) &=& \biggl[\left(-\frac{\partial}{\partial\tau} +
\frac{1}{2m_{\text{e}}}\,{\bm\nabla}^2 + \mu\right)\,\sigma_0
\nonumber\\
&& + \Gamma_{\text{t}}\,{\bm M}_{\text{sp}}({\bm
x})\cdot{\bm\sigma}\biggr]\,\delta(x-y).
\label{eq:4.12a}
\eea
Here ${\bm\sigma} = \left(\sigma_1,\sigma_2,\sigma_3\right)$ denotes the Pauli
matrices, and $\sigma_0$ is the $2\times 2$ unit matrix. Upon Fourier
transformation we have
\be
G^{-1}_{{\bm k},{\bm p}}(i\omega_n) = \delta_{{\bm k},{\bm p}}\,G_0^{-1}({\bm
k},i\omega_n)\,\sigma_0 + \Gamma_{\text{t}}\,{\bm M}_{\text{sp}}({\bm k}-{\bm
p})\cdot{\bm\sigma}.
\label{eq:4.12b}
\ee
\ese
The result of the inversion problem is
\bse
\label{eqs:4.13}
\bea
G_{{\bm k}{\bm p}}(i\omega_n) &=& \delta_{{\bm k}{\bm
p}}\,\bigl[\sigma_{+-}\,a_+({\bm k},{\bm q};i\omega_n) + \sigma_{-+}\,a_-({\bm
k},{\bm q};i\omega_n)\bigr]
\nonumber\\
&&\hskip 20pt + \delta_{{\bm k}+{\bm q},{\bm p}}\,\sigma_+\,b_+({\bm k},{\bm
q};i\omega_n)
\nonumber\\
&&\hskip 20pt   + \delta_{{\bm k}-{\bm q},{\bm p}}\,\sigma_-\,b_-({\bm k},{\bm
q};i\omega_n),
\label{eq:4.13a}
\eea
where
\bea
a_{\pm}({\bm k},{\bm q};i\omega_n) = \frac{G_0^{-1}({\bm k}\pm{\bm
q},i\omega_n)}{G_0^{-1}({\bm k},i\omega_n)\,G_0^{-1}({\bm k}\pm{\bm
q},i\omega_n) - \lambda^2}\ ,
\nonumber\\
\label{eq:4.13b}\\
b_{\pm}({\bm k},{\bm q};i\omega_n) = \frac{-\lambda}{G_0^{-1}({\bm
k},i\omega_n)\,G_0^{-1}({\bm k}\pm{\bm q},i\omega_n) - \lambda^2}\ .
\nonumber\\
\label{eq:4.13c}
\eea
\ese
Here $\sigma_{\pm} = (\sigma_1 \pm i\sigma_2)/2$, $\sigma_{+-} =
\sigma_+\sigma_-$, and $\sigma_{-+} = \sigma_-\sigma_+$.

\subsubsection{Spin susceptibility}
\label{subsubsec:IV.B.3}

Since the reference ensemble describes noninteracting electrons, the reference
ensemble spin susceptibility factorizes into a product of two Green functions.
Applying Wick's theorem to Eq.\ (\ref{eq:4.8d}) one obtains
\bse
\label{eqs:4.14}
\be
\chi_0^{ij}(x,y) = -\tr \bigl(\sigma_i\,G(x,y)\,\sigma_j\,G(y,x)\bigr),
\label{eq:4.14a}
\ee
or, after a Fourier transform,
\bea
\chi_0^{ij}({\bm k},{\bm p};i\Omega_n) &=& \frac{-1}{V}\sum_{{\bm k}',{\bm p}'}
T\sum_{i\omega_n} \tr \bigl( \sigma_i\,G_{{\bm k}',{\bm p}'}(i\omega_n)\,
\nonumber\\
&&\hskip -30pt \times \sigma_j\,G_{{\bm p}'+{\bm p},{\bm k}'+{\bm k}}(i\omega_n
+ i\Omega_n)\bigr).
\label{eq:4.14b}
\eea
\ese
Here the trace is over the spin degrees of freedom, and $\Omega_n = 2\pi Tn$ is
a bosonic Matsubara frequency.

From the structure of the Green function, Eq.\ (\ref{eq:4.13a}), it is obvious
that $\chi_0$ is nonzero if ${\bm k}$ and ${\bm p}$ differ by zero, $\pm{\bm
q}$, or $\pm 2{\bm q}$. The full expression in terms of $a_{\pm}$ and $b_{\pm}$
is lengthy and given in Appendix\ \ref{app:A}.

\subsubsection{Ferromagnetic limit}
\label{subsubsec:IV.B.4}

It is illustrative to check the ferromagnetic limit, ${\bm q}\to 0$, at this
point. In this case ${\bm M}_{\text{sp}} = (m_0,0,0)$ becomes position
independent, and both the Green function and the reference ensemble spin
susceptibility become diagonal in momentum space. For zero momentum and
frequency, the latter is also diagonal in spin space,
\bse
\label{eqs:4.15}
\bea
\chi_{0,{\bm q}=0}^{ij}({\bm k},{\bm p};i\Omega_n) = \delta_{{\bm k},{\bm p}}\,
   \chi_{0,{\bm q}=0}^{ij}({\bm k},i\Omega_n),
\label{eq:4.15a}\\
\chi_{0,{\bm q}=0}^{ij}(0,i0) = \delta_{ij}\,\left[\delta_{i1}\,\chi_{\text{L}}
+ (1-\delta_{i1})\,\chi_{\text{T}}\right].
\label{eq:4.15b}
\eea
\ese
The static and homogeneous transverse susceptibility $\chi_{\text{T}}$ of the
reference ensemble is related to the magnetization by a Ward
identity\cite{Ma_1976_ch_IX.7, Zinn-Justin_1996_ch_13.4} (remember that
$\Gamma_{\text{t}}{\bm M}_{\text{sp}}$ is the effective field in the reference
ensemble, see Eq.\ (\ref{eq:4.7b}))
\be
\langle{\bm n}_{\text{s}}(x)\rangle_0 = \Gamma_{\text{t}}\,{\bm
M}_{\text{sp}}\,\chi_{\text{T}}.
\label{eq:4.16}
\ee
A calculation of $\chi_{\text{T}}$ by evaluating Eq.\ (\ref{eq:4.14b}) for
${\bm q}=0$ shows that Eq.\ (\ref{eq:4.16}) is the equation of state, Eq.\
(\ref{eq:4.10c}), for ${\bm q}=0$. Equation (\ref{eq:4.10c}) thus represents
the generalization of this Ward identity to the helimagnetic case.

\subsection{Gaussian fluctuations I: ${\bm k} = {\bm q}$ modes}
\label{subsec:IV.C}

We are now in a position to explicitly write down the fluctuation action given
by Eqs.\ (\ref{eqs:4.8}). From both the phenomenological arguments in Sec.\
\ref{subsec:II.A} and the classical field theory in Sec.\ \ref{subsec:III.A} we
expect the static behavior to be correctly described by the fluctuations with
wave numbers close to the pitch wave number $q$, while Sec.\ \ref{subsec:II.B}
suggests that treating the dynamics correctly requires to also take into
account fluctuations with wave numbers near zero. For the sake of transparency
we first concentrate on the ${\bm k} = {\bm q}$ modes. We will later expand our
set of modes to study the effects of the ${\bm k} = 0$ modes on the dynamics.

\subsubsection{Gaussian action}
\label{subsubsec:IV.C.1}

We parameterize the fluctuations of the order parameter as in the classical
case, Eqs.\ (\ref{eqs:3.4}), but now allow for the fields $\phi$, $\varphi_1$,
and $\varphi_2$ to depend on imaginary time or Matsubara frequency. To linear
order in the fluctuations we have
\bea
\delta{\bm M}(x) = m_0 \left(
     \begin{array}{c} -\phi(x)\,\sin ({\bm q}\cdot{\bm x}) \\
                       \phi(x)\,\cos ({\bm q}\cdot{\bm x}) \\
                       \varphi_1(x)\,\sin ({\bm q}\cdot{\bm x})
                                     + \varphi_2(x)\,\cos ({\bm q}\cdot{\bm x})
     \end{array} \right)
     \nonumber\\
\label{eq:4.17}
\eea
As in the classical case we have anticipated that fluctuations of the norm of
the order parameter are massive. The term linear in $\delta{\bm M}$ vanishes
due to the saddle-point condition, and the Gaussian term can be expressed in
terms of integrals by using Eqs.\ (\ref{eq:4.14b}) and
(\ref{eq:A.1},\ref{eqs:A.2}). Using the notation $\phi(x) \equiv \varphi_0(x)$
as in the classical case, we find a Gaussian action \cite{fm_limit_footnote}
\bse
\label{eqs:4.18}
\bea
{\cal A}^{(2)}[\varphi_i] &=& \frac{\lambda^2}{2} \sum_{\bm p}\sum_{i\Omega_n}
\sum_{i=0,1,2} \varphi_i({\bm p},i\Omega_n)\,\gamma^{({\bm q})}_{ij}({\bm
p},i\Omega_n)\,
\nonumber\\
&& \hskip 70pt \times \varphi_j(-{\bm p},-i\Omega_n).
\label{eq:4.18a}
\eea
Here the matrix $\gamma^{({\bm q})}$ is the quantum mechanical analog of Eq.\
(\ref{eq:3.5b}), which couples the phase or ${\bm k}={\bm q}$ modes among each
other. In a four-vector notation, $k \equiv ({\bm k},i\Omega_n) \equiv
(k_x,k_y,k_z,i\Omega_n)$ it is given by
\begin{widetext}
\be
\gamma^{({\bm q})}(k) = \left(\begin{array}{ccc}
    (1 - cq)/\Gamma_{\text{t}} - f_{\phi\phi}(k) &
                          -ick_y/2\Gamma_{\text{t}} & -ick_x/2\Gamma_{\text{t}} \\
    ick_y/2\Gamma_{\text{t}} & 1/2\Gamma_{\text{t}} - f_{11}(k)
                                                    & -f_{12}(k) \\
    ick_x/2\Gamma_{\text{t}} & f_{12}(k) & 1/2\Gamma_{\text{t}}
                                                    - f_{11}(k)
    \end{array} \right).
\label{eq:4.18b}
\ee
Here
\bea
f_{\phi\phi}(k) &=& \varphi_{\phi\phi}(k) +
                    \varphi_{\phi\phi}(-k),
\label{eq:4.18c}\\
f_{11}(k) &=& \varphi_{11}(k) +
                    \varphi_{11}(-k),
\label{eq:4.18d}\\
f_{12}(k) &=& i\bigl[\varphi_{11}(k) -
                    \varphi_{11}(-k)\bigr],
\label{eq:4.18e}
\eea
where
\bea
\varphi_{\phi\phi}(k) &=& -\int_p \frac{G_0^{-1}({\bm p}-{\bm k},i\omega_m -
i\Omega_n) G_0^{-1}({\bm p}-{\bm q},i\omega_m) - \lambda^2}{\bigl[G_0^{-1}({\bm
p}-{\bm k},i\omega_m-i\Omega_n)\,G_0^{-1}({\bm p}-{\bm k}-{\bm
q},i\omega_m-i\Omega_n) - \lambda^2\bigr]\,\bigl[G_0^{-1}({\bm
p},i\omega_m)\,G_0^{-1}({\bm p}-{\bm q},i\omega_m) - \lambda^2\bigr]},
\nonumber\\
\label{eq:4.18f}\\
\varphi_{11}(k) &=& \frac{-1}{4}\int_p \frac{G_0^{-1}({\bm p}-{\bm k},i\omega_m
- i\Omega_n) G_0^{-1}({\bm p}+{\bm q},i\omega_m) + G_0^{-1}({\bm p}-{\bm
k}-{\bm q},i\omega_m-i\Omega_n)\,G_0^{-1}({\bm p},i\omega_m) -
2\lambda^2}{\bigl[G_0^{-1}({\bm p}-{\bm k},i\omega_m-i\Omega_n)\,G_0^{-1}({\bm
p}-{\bm k}-{\bm q},i\omega_m-i\Omega_n) - \lambda^2\bigr]\,\bigl[G_0^{-1}({\bm
p},i\omega_m)\,G_0^{-1}({\bm p}+{\bm q},i\omega_m) - \lambda^2\bigr]},
\nonumber\\
\label{eq:4.18g}
\eea
\end{widetext}
\ese
with $\int_p \equiv (1/V)\sum_{\bm p} T\sum_{i\omega_m}$.

In contrast to the classical case, here it is not obvious that the Gaussian
vertex, Eq.\ (\ref{eq:4.18b}) has a zero eigenvalue. To see that it does, we
invoke the equation of state (\ref{eq:4.10c}). By comparing this with Eqs.\
(\ref{eq:4.18c}, \ref{eq:4.18f}), we see that
\bse
\label{eqs:4.19}
\be
1 - c\,q - \Gamma_{\text{t}} f_{\phi\phi}(0,i0) = 0.
\label{eq:4.19a}
\ee
Similarly,
\be
1/2 - \Gamma_{\text{t}} f_{11}(0,i0) = c\,q/2.
\label{eq:4.19b}
\ee
\ese
Since $c\propto q$ (see Eqs.\ (\ref{eq:3.2a}) and (\ref{eq:4.21}) below), it
follows that the quantum mechanical vertex $\gamma^{(\bm q)}({\bm
p},i\Omega_n)$ has the same structure as its classical counterpart, Eq.\
(\ref{eq:3.5b}), except for an additional frequency dependence in the quantum
mechanical case.

To determine the eigenvalues we need to evaluate the integrals to lowest
nontrivial order in the wave vector and the frequency. A complete calculation
is rather difficult, and we restrict ourselves to the limit $\lambda \gg
qv_{\text{F}}$. The calculation, the details of which we relegate to Appendix
{\ref{app:B}, yields
\bse
\label{eqs:4.20}
\bea
f_{\phi\phi}({\bm k},i\Omega_n) &=& f_{\phi\phi}(0,i0)
-2N_{\text{F}}\biggl[\alpha\left(\frac{{\bm k}}{2k_{\text{F}}}\right)^2 \hskip
30pt
\nonumber\\
&& \hskip -30pt - \beta\left(\frac{i\Omega_n}{4\epsilon_{\text{F}}}\right)^2 +
\gamma_{\phi}\,
\frac{\vert\Omega_n\vert}{4\epsilon_{\text{F}}}\,\frac{k_z^2}{2k_{\text{F}}\vert{\bm
k}\vert}\biggr]\ ,
\label{eq:4.20a}
\eea
\bea
f_{11}({\bm k},i\Omega_n) &=& f_{11}(0,i0) -
N_{\text{F}}\biggl[\alpha\left(\frac{{\bm k}}{2k_{\text{F}}}\right)^2 \hskip
30pt
\nonumber\\
&& \hskip -30pt - \beta\left(\frac{i\Omega_n}{4\epsilon_{\text{F}}}\right)^2 +
\gamma_{1}\,
\frac{\vert\Omega_n\vert}{4\epsilon_{\text{F}}}\,
               \frac{{\bm k}_{\perp}^2}{(2k_{\text{F}})^2}\biggr]\
,
\label{eq:4.20b}\\
f_{12}({\bm k},i\Omega_n) &=& -i2N_{\text{F}}\,\alpha\,\frac{q
k_z}{(2k_{\text{F}})^2}
\label{eq:4.20c}
\eea
Here
\bea
\alpha &=& 1/3,
\label{eq:4.20d}\\
\beta &=& 4\epsilon_{\text{F}}^2/\lambda^2,
\label{eq:4.20e}\\
\gamma_{\phi} &=& \pi (qv_{\text{F}})^2/8\lambda^2,
\label{eq:4.20f}\\
\gamma_1 &=& -4\gamma_{\phi}(k_{\text{F}}/q)^3.
\label{eq:4.20g}
\eea
\ese
These expressions are valid for $\vert\Omega_n\vert \ll \lambda \ll
\epsilon_{\text{F}}$, $\vert{\bm k}\vert \ll q \ll k_{\text{F}}$, and
$qv_{\text{F}} \ll \lambda$. The damping terms have the form shown if, in
addition, $\vert\Omega\vert \ll v_{\text{F}}\vert{\bm k}\vert$.

We now also can express the pitch wave number $q$ in terms of the parameters of
our model. The minimization of the saddle-point free energy proceeds as in the
classical case, and by comparing Eq. (\ref{eq:3.2a}) with Eqs. (\ref{eq:4.17},
\ref{eq:4.18a}, \ref{eq:4.18b}, \ref{eq:4.19a}, \ref{eq:4.20a}), we find
\be
q = c\,k_{\text{F}}^2/N_{\text{F}}\Gamma_{\text{t}}\alpha.
\label{eq:4.21}
\ee

\subsubsection{Eigenvalue problem}
\label{subsubsec:IV.C.2}

The diagonalization of the matrix $\gamma^{({\bm q})}$, Eq.\ (\ref{eq:4.18b}),
can be done perturbatively as in the classical case. The soft (Goldstone) mode,
i.e., the eigenvector corresponding to the smallest eigenvalue, is
\bea
v({\bm k},i\Omega_n) &=& \phi({\bm k},i\Omega_n) - i(k_y/q)[1 + O({\bm
k}_{\perp}^2)]\,\varphi_1({\bm k},i\Omega_n)
\nonumber\\
&& - i(k_x/q)[1 + O({\bm k}_{\perp}^2)]\,\varphi_2({\bm k},i\Omega_n).
\label{eq:4.22}
\eea
The helimagnon is proportional to the $v$-$v$-correlation function, which in
turn is proportional to the inverse of the smallest eigenvalue of the matrix
$\gamma^{({\bm q})}$. With ${\bm\kappa} = \sqrt{\alpha}{\bm k}/2k_{\text{F}}$,
${\bm Q} = \sqrt{\alpha}{\bm q}/2k_{\text{F}}$, $\omega =
\sqrt{\beta}\Omega_n/4\epsilon_{\text{F}} \equiv \Omega_n/2\lambda$, $c_{\phi}
= \gamma_{\phi}/\sqrt{\alpha\beta}$, and $c_1 = \gamma_1/\alpha\sqrt{\beta}$,
the latter reads
\begin{widetext}
\be
\gamma^{({\bm q})}(k) = 2N_{\text{F}} \left(\begin{array}{ccc}
   \kappa^2 + \omega^2 +
      c_{\phi}\vert\omega\vert\,\kappa_z^2/\vert{\bm\kappa}\vert &
                                   -i\,Q\,\kappa_y & -i\,Q\,\kappa_x \cr
   i\,Q\,\kappa_y & Q^2 +
      \frac{1}{2}\,\kappa^2 + \frac{1}{2}\,\omega^2 -
      \frac{1}{2}\,c_1\vert\omega\vert\,{\bm\kappa}_{\perp}^2 & i\,Q\,\kappa_z \cr
   i\,Q\,\kappa_x & -i\,Q\,\kappa_z & Q^2 +
      \frac{1}{2}\,\kappa^2 + \frac{1}{2}\,\omega^2 -
      \frac{1}{2}\,c_1\vert\omega\vert\, {\bm\kappa}_{\perp}^2
                      \end{array}\right)
\label{eq:4.23}
\ee
\end{widetext}

The eigenvalue equation, which is the quantum mechanical generalization of Eq.\
(\ref{eq:3.6}), can be simplified if we anticipate, from Secs.\ \ref{sec:II}
and \ref{subsec:III.B}, that the smallest eigenvalue scales as $\mu \sim
\kappa_z^2 \sim {\bm\kappa}_{\perp}^4 \sim \omega^2$. Keeping only terms up to
$O(\kappa_z^2)$, the eigenvalue equation reads
\bea
\left(\mu - \kappa^2 - \omega^2 -
c_{\phi}\,\vert\omega\vert\,\frac{\kappa_z^2}{\vert{\bm\kappa}\vert}\right)
\left(Q^4 + Q^2\kappa_{\perp}^2\right)
\nonumber\\
+\,Q^2\,{\bm\kappa}_{\perp}^2\left(Q^2 + \frac{{\bm\kappa}_{\perp}^2}{2}\right)
= 0.
\label{eq:4.24}
\eea

This result has several interesting aspects, which will be very useful when we
generalize the theory to include the ${\bm k}=0$ modes in Sec.\
\ref{subsec:IV.D}. First, of the function $f_{11}$, Eq.\ (\ref{eq:4.20b}), only
the constant and the term proportional to ${\bm k}_{\perp}^2 \sim k_z$
contribute to the leading terms in the eigenvalue. In particular, the damping
term proportional to $\gamma_1$, which has a potential to lead to an
overdamping of the Goldstone mode, does not contribute. Second, the function
$f_{12}$, which describes the coupling between the phase modes $\varphi_1$ and
$\varphi_2$, does not contribute to the leading result. Indeed, the only role
of $\varphi_1$ and $\varphi_2$ is to subtract the ${\bm\kappa}_{\perp}^2$
contribution from the eigenvalue, and this is accomplished entirely by the
purely static matrix elements in Eq.\ (\ref{eq:4.18b}) which do not depend on
reference ensemble correlation functions.

For the smallest eigenvalue we find from Eq.\ (\ref{eq:4.24})
\be
\mu({\bm\kappa}\to 0,\omega\to 0) = \kappa_z^2 + \omega^2 +
\frac{{\bm\kappa}_{\perp}^4}{2Q^2} +
c_{\phi}\,\vert\omega\vert\,\frac{\kappa_z^2}{\vert{\bm\kappa}\vert} +
O(\kappa_z^3).
\label{eq:4.25}
\ee

Since the Goldstone correlation function is proportional to the inverse of the
smallest eigenvalue, this result has indeed the functional form we expect from
the phenomenological treatment in Sec.\ \ref{sec:II}, see Eqs.\
(\ref{eq:2.18a}, \ref{eq:2.19a}). At zero frequency it also is consistent with
the result of the classical field theory in Sec.\ \ref{sec:III}. However, in
contrast to Eq.\ (\ref{eq:2.13}) the mass of the zero-wave number mode does not
enter in Eq.\ (\ref{eq:4.25}), which suggests that the frequency scale is not
correctly described yet. This was to be expected, see the remarks at the start
of Sec.\ \ref{subsec:IV.C}. We will see in Sec.\ \ref{subsec:IV.D} that the
${\bm k}=0$ mode has to be included in the analysis to obtain the correct
prefactor of the $\omega^2$ term in the smallest eigenvalue, in agreement with
the phenomenological analysis.

\subsubsection{Ferromagnetic limit}
\label{subsubsec:IV.C.3}

Before we expand our set of modes it is again illustrative to consider the
ferromagnetic limit ${\bm q}=0$. Going back to Eqs.\ (\ref{eq:4.17}), we see
that in this limit $\varphi_1$ disappears, and $\phi$ and $\varphi_2$ play the
roles of the two independent field components $\pi_1$ and $\pi_2$ in a
ferromagnetic nonlinear sigma model, which are both
soft.\cite{Zinn-Justin_1996_ch_30} The Gaussian action is given by the
reference ensemble spin susceptibility at ${\bm q}=0$, and one
finds\cite{Belitz_et_al_1998}
\bea
{\cal A}^{(2)}[\phi,\varphi_2] &=&
\frac{1}{2}\,N_{\text{F}}\,\Gamma_{\text{t}}^2
   \sum_{{\bm p},i\Omega_n}\sum_{i=1,2}\pi_i({\bm
   p},i\Omega_n)\,{\tilde\gamma}_{ij}({\bm p},i\Omega_n)
\nonumber\\
&&\hskip 50pt \times \pi_j(-{\bm p},-i\Omega_n),
\label{eq:4.26}
\eea
with $\pi_1\equiv\phi$ and $\pi_2\equiv\varphi_2$, and a matrix
\bea
{\tilde\gamma}_{ij}({\bm p},i\Omega_n) = \left(\begin{array}{cc}
    \kappa^2   &   i(i\omega)  \cr
    -i(i\omega)  &  \kappa^2   \end{array}\right).
\label{eq:4.27}
\eea
Comparing with the eigenvalue problem given by Eq.\ (\ref{eq:4.23}) we see that
the latter does {\em not} correctly reproduce the ferromagnetic result upon
dropping $\varphi_1$ and letting ${\bm q}\to 0$. This is not surprising, since
the gradient expansion implicit in the effective field theory approach implies
that we are restricted to wave numbers small compared to $q$. That is, the
validity of Eq.\ (\ref{eq:4.23}) shrinks to zero as ${\bm q}\to 0$. However, it
raises the following question. The leading frequency dependence of the
ferromagnetic magnon is determined by the off-diagonal elements of the matrix
${\tilde\gamma}$. Similarly, the time-dependent Ginzburg-Landau theory of Sec.\
\ref{subsec:II.B} suggests that the leading frequency dependence of the
helimagnon is produced by the coupling of the phase mode to the homogeneous
magnetization. Although the latter is massive at wave number $q$, its conserved
character makes it important for the dynamics. In contrast, in the treatment
above the leading frequency dependence was produced by the phase correlation
functions. We therefore extend our set of modes to allow for a coupling between
the phase modes, which represent ${\bm k} = {\bm q}$ fluctuations of the
magnetization, and the homogeneous magnetization.

\subsection{Gaussian fluctuations II: coupling between ${\bm k}={\bm q}$ modes
            and ${\bm k}=0$ modes}
\label{subsec:IV.D}

The discussion in the preceding subsection suggests to generalize the
expression for the magnetization fluctuations, Eq.\ (\ref{eq:4.17}), by writing
\bea
\delta{\bm M}(x) = m_0 \left(
     \begin{array}{c} -\phi(x)\,\sin ({\bm q}\cdot{\bm x}) \\
                       \phi(x)\,\cos ({\bm q}\cdot{\bm x}) + \pi_2(x) \\
                       \varphi_1(x)\,\sin ({\bm q}\cdot{\bm x})
                                     + \varphi_2(x)\,\cos ({\bm q}\cdot{\bm x}) \\
                                     \hskip 95pt + \pi_1(x)
     \end{array} \right).
\nonumber\\
\label{eq:4.28}
\eea
That is, we add the ${\bm k}=0$ fluctuations to the ${\bm k}={\bm q}$
fluctuations.\cite{mode_coupling_footnote}

\subsubsection{Gaussian action}
\label{subsubsec:IV.D.1}

If one repeats the development of the previous subsection for the current set
of five modes, one finds that of the two ${\bm k}=0$ modes only $\pi_1$
contributes to the leading terms in the eigenvalue problem. $\pi_2$ does not
couple to $\phi$, and its couplings to $\varphi_1$ and $\varphi_2$ produce only
higher order corrections. $\pi_1$, on the other hand, couples to $\phi$ in a
way that preserves the off-diagonal frequency terms characteristic for the
ferromagnetic problem, see Eq.\ (\ref{eq:4.27}), and needs to be kept. These
observations are in agreement with the phenomenological theory of Sec.\
\ref{subsec:II.B}, where only the 3-component of the homogeneous magnetization
coupled to the phase mode. They also are consistent with the observation in
Sec.\ \ref{subsubsec:IV.C.2} that $\varphi_1$ and $\varphi_2$ serve only to
provide the correct static structure of the theory. We thus drop $\pi_2$ and
consider the $4\times 4$ problem given by the phase modes plus $\pi_1$.

The Gaussian action that generalizes Eqs.\ (\ref{eqs:4.18}) now reads
\bse
\label{eqs:4.29}
\bea
{\cal A}^{(2)}[\varphi_i] &=& \frac{\lambda^2}{2} \sum_{\bm p}\sum_{i\Omega_n}
\sum_{i=0}^{3} \varphi_i({\bm p},i\Omega_n)\,\gamma^{({\bm q},0)}_{ij}({\bm
p},i\Omega_n)\,
\nonumber\\
&& \hskip 70pt \times \varphi_j(-{\bm p},-i\Omega_n).
\label{eq:4.29a}
\\\nonumber
\eea
Here $\varphi_3 \equiv \pi_1$, and the matrix $\gamma^{({\bm q},0)}$, which
couples the phase modes to $\pi_1$, reads, in a block matrix notation,
\bea
\gamma^{({\bm q},0)}(k) = \left(\begin{array}{ccccc}
                                    & \bigl\vert & -ih_{\phi 1}(k)   \cr
              \gamma^{({\bm q})}(k) & \bigl\vert &   0               \cr
                                    & \bigl\vert &   0               \cr
   \text{\leaders\hrule\hskip 60pt} & \bigl\vert & \text{\leaders\hrule\hskip
60pt} \cr
    ih_{\phi 1}(k)\ \ 0\ \ 0  & \bigl\vert & 1/\Gamma_{\text{t}} -
    g_{11}(k)
    \end{array} \right)
    \nonumber\\
\label{eq:4.29b}
\eea
with $\gamma^{({\bm q})}$ from Eq.\ (\ref{eq:4.18b}). In addition to the
functions defined in Eqs.\ (\ref{eqs:4.18}), we need
\be
g_{11}({\bm k},i\Omega_n) = 4\varphi_{11}({\bm k}-{\bm q},i\Omega_n),
\label{eq:4.29c}
\ee
with $\varphi_{11}$ from Eq.\ (\ref{eq:4.18g}), and
\be
h_{\phi 1}({\bm k},i\Omega_n) = \eta_{\phi 1}({\bm k},i\Omega_n)
                                - \eta_{\phi 1}(-{\bm k},-i\Omega_n),
\label{eq:4.29d}
\ee
where
\begin{widetext}
\bea
\eta_{\phi 1}(k) = \lambda\int_p \frac{G_0^{-1}({\bm p}-{\bm k},i\omega_m -
i\Omega_n) - G_0^{-1}({\bm p}-{\bm q},i\omega_m)}{\bigl[G_0^{-1}({\bm p}-{\bm
k},i\omega_m-i\Omega_n)\,G_0^{-1}({\bm p}-{\bm k}-{\bm q},i\omega_m-i\Omega_n)
- \lambda^2\bigr]\,\bigl[G_0^{-1}({\bm p},i\omega_m)\,G_0^{-1}({\bm p}-{\bm
q},i\omega_m) - \lambda^2\bigr]}.
\nonumber\\
\label{eq:4.29e}
\eea
\end{widetext}
\ese
Performing these integrals in the limit $\lambda \gg qv_{\text{F}}$ yields the
following leading contributions (see Appendix \ref{app:B}),
\bse
\label{eqs:4.30}
\bea
g_{11}({\bm k},i\Omega_n) &=& 1/\Gamma_{\text{t}} - 2N_{\text{F}}\Bigl[ \alpha
\left(\frac{{\bm q}}{2k_{\text{F}}}\right)^2 + \alpha \left(\frac{{\bm
k}}{2k_{\text{F}}}\right)^2
\nonumber\\
&& \hskip -20pt + \beta\left(\frac{i\Omega_n}{4\epsilon_{\text{F}}}\right)^2 -
\frac{1}{2}\,\gamma_{\phi}\,\frac{\vert\Omega_n\vert}{4\epsilon_{\text{F}}}\,
\frac{2k_{\text{F}}{\bm k}_{\perp}^2}{\vert{\bm k}\vert^3}\Bigr].
\label{eq:4.30a}\\
h_{\phi 1}({\bm k},i\Omega_n) &=&
-2N_{\text{F}}\,\sqrt{\beta}\,\frac{i\Omega_n}{4\epsilon_{\text{F}}}.
\label{eq:4.30b}
\eea
\ese

\subsubsection{Eigenvalue problem}
\label{subsubsec:IV.D.2}

Now consider the generalization of Eq.\ (\ref{eq:4.23}). It is obvious from
Eqs.\ (\ref{eq:4.29b}, \ref{eqs:4.30}) that the mode $\pi_1$ will contribute to
the eigenvalue a term proportional to $\Omega^2$ whose prefactor is large
compared to the one in Eq.\ (\ref{eq:4.25}) by a factor of
$(k_{\text{F}}/q)^2$. This is because the mass of $\pi_1$ is proportional to
$(q/k_{\text{F}})^2$. Consequently, one can neglect the terms proportional to
$\omega^2$ in the diagonal elements of $\gamma^{({\bm q})}$, Eq.\
(\ref{eq:4.23}). The leading damping term, however, is still given by
$f_{\phi\phi}$, Eq.\ (\ref{eq:4.20a}). Dropping all terms that yield
contributions of higher order than $k_z^2$ in the eigenvalue equation, the
matrix $\gamma^{({\bm q},0)}$, Eq.\ (\ref{eq:4.29b}), reads, in the notation of
Sec.\ \ref{subsubsec:IV.C.2},
\bea
\gamma^{({\bm q},0)}(k) &=& \nonumber\\
&& \hskip -50pt 2N_{\text{F}}\left(\begin{array}{cccc}
     \kappa^2 + c_{\phi}\vert\omega\vert\,\frac{\kappa_z^2}{\vert{\bm\kappa}\vert} &
          -i Q \kappa_y   &   -i Q \kappa_x   &   i(i\omega)   \cr
     i Q \kappa_y   &   Q^2 + \frac{1}{2}\,{\bm\kappa}_{\perp}^2   &   i Q \kappa_z
           &   0   \cr
     i Q \kappa_x   &   -i Q \kappa_z   &   Q^2 + \frac{1}{2}\,{\bm\kappa}_{\perp}^2
           &   0   \cr
     -i(i\omega)   &   0   &   0   &   Q^2
     \end{array}\right).
     \nonumber\\
\label{eq:4.31}
\eea
The smallest eigenvalue is
\bse
\label{eqs:4.32}
\be
\mu({\bm\kappa}\to 0,\omega\to 0) = \kappa_z^2 + \frac{\omega^2}{Q^2} +
\frac{{\bm\kappa}_{\perp}^4}{2Q^2} +
c_{\phi}\,\vert\omega\vert\,\frac{\kappa_z^2}{\vert{\bm\kappa}\vert} +
O(\kappa_z^3),
\label{eq:4.32a}
\ee
and the corresponding eigenvector reads
\bea
v({\bm k},i\Omega_n) &=& \phi({\bm k},i\Omega_n) -
i(\kappa_y/Q)\,\varphi_1({\bm k},i\Omega_n)
\nonumber\\
&& -i(\kappa_x/Q)\,\varphi_2({\bm k},i\Omega_n) + (4\omega/Q^2)\,\pi_1({\bm
k},i\Omega_n).
\nonumber\\
\label{eq:4.32b}
\eea
\ese

Notice that the origin of the leading frequency dependence is consistent with
the ferromagnetic limit, Eq.\ (\ref{eq:4.27}), and, more importantly, with the
time-dependent Ginzburg-Landau theory in Sec.\ \ref{subsec:II.B}. Namely, it is
produced by the coupling of the $3$-component of the ${\bm k}=0$ magnetization
fluctuation to the phase mode at ${\bm k}={\bm q}$. The frequency dependence in
Eq.\ (\ref{eq:4.23}), on the other hand, corresponded to second order time
derivative corrections to the kinetic equation (\ref{eq:2.6}).

\subsection{The Goldstone mode, and the spin susceptibility}
\label{subsec:IV.E}

\subsubsection{The Goldstone mode in the clean limit}
\label{subsubsec:IV.E.1}

We are now in a position to determine the Goldstone mode. Defining the latter
as $g({\bm k},i\Omega_n) = (\sqrt{2N_{\text{F}}}\sqrt{3}k_{\text{F}}/q)\,v({\bm
k},i\Omega_n)$, and returning to ordinary units, we find from Eqs.\
(\ref{eqs:4.32})
\bse
\label{eqs:4.33}
\be
\left\langle g({\bm k},i\Omega_n)\,g(-{\bm k},-i\Omega_n)\right\rangle =
\frac{1}{-(i\Omega_n)^2 + \omega_0^2({\bm k}) + \vert\Omega_n\vert\gamma({\bm
k})},
\label{eq:4.33a}
\ee
where
\be
\omega_0({\bm k}) =
\lambda\,\frac{q}{3k_{\text{F}}}\,\sqrt{k_z^2/(2k_{\text{F}})^2 +
\frac{1}{2}\,{\bm k}_{\perp}^4/(2qk_{\text{F}})^2},
\label{eq:4.33b}
\ee
and
\be
\gamma({\bm k}) = \frac{2\pi}{3}\,\epsilon_{\text{F}}\,
\frac{q^4}{(2k_{\text{F}})^4}\, \frac{k_z^2}{2k_{\text{F}}\vert{\bm k}\vert}
\label{eq:4.33c}
\ee
\ese
This has the same functional form as the result of the phenomenological
treatment in Sec.\ \ref{sec:II}, see Eqs.\ (\ref{eq:2.18a}, \ref{eq:2.19a}).
The current microscopic derivation reveals in addition that the prefactor of
the damping coefficient is smaller than that of the resonance frequency by at
least a factor of $(qv_{\text{F}}/\lambda)(q/k_{\text{F}})$. The Goldstone mode
is thus propagating and weakly damped for all orientations of the wave
vector.\cite{damping_footnote} As in the classical case, adding cubic
anisotropic terms, e.g., the quantum mechanical generalization of Eq.\
(\ref{eq:2.21}), leads to a soft-mode energy of the form of Eq.\
(\ref{eq:2.23}).

\subsubsection{The Goldstone mode in the presence of quenched disorder}
\label{subsubsec:IV.E.2}

Equation (\ref{eq:4.33c}) holds for clean systems. As we have mentioned in the
context of Eqs.\ (\ref{eqs:2.19}), the structure of the damping term is
expected to change qualitatively in the presence of even weak impurity
scattering. Let $\tau_{\text{\,imp}}$ be the inelastic scattering rate due to
quenched impurities. Then the bare Green function $G_0$, Eq.\ (\ref{eq:4.9a}),
acquires a finite lifetime,\cite{AGD_1963}
\be
G_0({\bm k},i\omega_n) = 1/(i\omega_n - \xi_{\bm k} +
(i/2\tau_{\text{\,imp}})\sgn \omega_n).
\label{eq:4.34}
\ee
For weak disorder, $\epsilon_{\text{F}}\tau_{\text{\,imp}} \gg 1$, the leading
effect is that the hydrodynamic singularity in the generalized Lindhard
function $\phi_{\phi\phi}^{(2)}$, Eq.\ (\ref{eq:B.1f}), is now protected. The
net effect is the replacement $1/v_{\text{F}}\vert{\bm k}\vert \to
\tau_{\text{\,imp}}$ in the damping term in $f_{\phi\phi}$, Eq.\
(\ref{eq:4.20a}), and hence in the damping coefficient $\gamma({\bm k})$. The
Goldstone mode is then given by Eq.\ (\ref{eq:4.33a}), with the resonance
frequency still given by Eq.\ (\ref{eq:4.33b}), and
\be
\gamma({\bm k}) = \frac{8\pi}{3}\,\epsilon_{\text{F}}\,
(\epsilon_{\text{F}}\tau_{\text{\,imp}})\, \frac{q^4}{(2k_{\text{F}})^4}\,
\frac{k_z^2}{(2k_{\text{F}})^2}.
\label{eq:4.35}
\ee
This is again consistent with the result of the phenomenological treatment in
Sec.\ \ref{sec:II}, see Eq.\ (\ref{eq:2.19b}). Since $\gamma({\bm k}) \sim
k_z^2 \ll \omega_0({\bm k}) \sim k_z$, the mode is again propagating.

\subsubsection{The physical spin susceptibility}
\label{subsubsec:IV.E.3}

The helimagnon correlation function is simply related to the physical
susceptibility, which is directly measurable by inelastic neutron
scattering.\cite{Shirane_et_al_1983, Ishida_et_al_1985, Roessli_et_al_2004} To
express the latter in terms of order-parameter correlation functions we
generalize the partition function, Eq.\ (\ref{eq:4.1a}), to a generating
functional for spin density correlation functions,
\be
Z[{\bm j}] = \int D[{\bar\psi},\psi]\,e^{S[{\bar\psi},\psi] + \int dx\,{\bm
j}(x)\cdot{\bm n}_{\text{s}}(x)},
\label{eq:4.36}
\ee
where ${\bm j}(x)$ is a source field. The spin susceptibility is given by
\bea
\chi_{\text{s}}^{ij}(x,y) &=& \langle n_{\text{s}}^i(x)\,
n_{\text{s}}^j(y)\rangle_S - \langle n_{\text{s}}^i(x)\rangle_S\,\langle
n_{\text{s}}^j(y)\rangle_S
\nonumber\\
&=& \frac{\delta^2}{\delta j_i(x)\,\delta j_j(y)}\,\biggr\vert_{{\bm j}=0}\ \ln
Z[{\bm j}]
\nonumber\\
&=& \Gamma_{\text{t}}^2\left\langle\left(A^{-1}\delta
M\right)_i(x)\,\left(A^{-1}\delta M\right)_j(y)\right\rangle_{\cal{A}}.
\nonumber\\
\label{eq:4.37}
\eea
Here $A^{-1}$ is the matrix given in Eq.\ (\ref{eq:4.2d}), and the last average
is to be taken with respect to the effective action ${\cal A}$, Eqs.\
(\ref{eqs:4.8}).

The chiral nature of the helix is also reflected in the spin fluctuations and
will become manifest if the spin susceptibility is measured with circularly
polarized neutrons. To see this, it is useful to define magnetization
fluctuations $\delta M_{\pm} = \delta M_1 \pm i\delta M_2$. Similarly, we
define gradient operators $\partial_{\pm} = \partial_x \pm i\partial_y$. With
these definitions one finds
\bse
\label{eqs:4.38}
\bea
\Gamma_{\text{t}}\left(A^{-1} \delta M\right)_{\pm}(x) &=& \delta M_{\pm}(x)
\pm i\,c\,\partial_z\,\delta M_{\pm}(x)
\nonumber\\
&& \mp i\,c\,\partial_{\pm}\,\delta M_z(x),
\label{eq:4.38a}\\
\Gamma_{\text{t}}\left(A^{-1} \delta M\right)_{z}(x) &=& \delta M_z(x) +
i\,\frac{c}{2}\,\bigl(\partial_+ \delta M_-(x)
\nonumber\\
&& \hskip 30pt - \partial_- \delta M_+(x)\bigr).
\label{eq:4.38b}
\eea
\ese
By means of Eq.\ (\ref{eq:4.28}) we can express the components of $\delta{\bm
M}$ in terms of $\varphi_0 \equiv \phi$, $\varphi_1$, $\varphi_2$, and
$\varphi_3\equiv \pi_1$,
\bse
\label{eqs:4.39}
\bea
\delta M_{\pm}(x) &=& \pm i\,m_0\,\phi(x)\,e^{\pm i{\bm q}\cdot{\bm x}},
\label{eq:4.39a}\\
\delta M_z(x) &=& m_0\,\pi_1(x) + \frac{m_0}{2}\,\bigl[\left(\varphi_2(x) -
i\varphi_1(x)\right)e^{\pm i{\bm q}\cdot{\bm x}}
\nonumber\\
&& \hskip 90pt + {\text{c.c}}\bigr],
\label{eq:4.39b}
\eea
\ese
where c.c. denotes the complex conjugate of the preceding expression. The
elements of the susceptibility tensor can therefore be expressed in terms of
the correlation functions of the $\varphi_i$, which we denote by
$\chi_{\phi\phi}$ etc. The various correlation functions can be determined from
the Gaussian action ${\cal A}^{(2)}$, Eq.\ (\ref{eq:4.29a}), with the vertex
function $\gamma^{({\bm q},0)}$ given by Eq.\ (\ref{eq:4.31}). The inverse of
the latter matrix reads
\begin{widetext}
\be
\left(\gamma^{({\bm q},0)}({\bm p},i\Omega_n)\right)^{-1} =
\frac{1}{2N_{\text{F}}\,Q^3\,\mu({\bm p},i\Omega_n)}
%\nonumber\\
 \left(\begin{array}{cccc}
      Q^3  &
      iQ^2\kappa_y  &
      iQ^2\kappa_x  &
      Q\omega  \cr
      -iQ^2\kappa_y &
      Q\kappa_y^2  &
      Q\kappa_x\kappa_y  &
      -i\omega\kappa_y   \cr
     -iQ^2\kappa_x  &
      Q\kappa_x\kappa_y  &
      Q\kappa_x^2  &
      -i\omega\kappa_x  \cr
      -Q\omega  &
      -i\omega\kappa_y  &
      -i\omega\kappa_x  &
      Q\kappa_z^2 + \kappa_{\perp}^4/2Q + Qc_{\phi}\vert\omega\vert\kappa_z^2/\vert{\bm\kappa}\vert
   \end{array}\right).
%\nonumber\\
\label{eq:4.40}
\ee
\end{widetext}
Here we use the same notation as is Eqs.\ (\ref{eq:4.23}) and (\ref{eq:4.31}),
and corrections to each matrix elements are one power higher in $\kappa_z \sim
{\bm\kappa}_{\perp}^2 \sim \omega$ than the terms shown. All of the correlation
functions that determine the spin susceptibility are proportional to the
inverse of the smallest eigenvalue $\mu$, Eq.\ (\ref{eq:4.32a}), which scales
as $\omega^2$. $\langle\phi\phi\rangle$ is the softest; it scales as
$1/\omega^2$. The autocorrelation functions of $\phi_1$ and $\phi_2$ have an
additional factor of ${\bm\kappa}_{\perp}^2 \sim \omega$ in the numerator, and
thus scale as $1/\omega$, and so does $\langle\phi_1\phi_2\rangle$. The
autocorrelation function of $\pi_1$ scales as a constant. The mixed
correlations $\langle\phi\phi_1\rangle$ and $\langle\phi\phi_2\rangle$ scale as
$1/\omega^{3/2}$, and the mixed correlations $\langle\phi\pi_1\rangle$ and
$\langle\phi_{1,2}\pi_1\rangle$ scale as $1/\omega$ and $1/\omega^{1/2}$,
respectively.

Defining the momentum and frequency dependent spin susceptibility by
\bea
\chi_{\text{s}}^{ij}({\bm k},{\bm p};i\Omega_n) &=& \int d{\bm x}\,d{\bm y}\
e^{i{\bm k}\cdot{\bm x} -i{\bm p}\cdot{\bm y}} \int_0^{1/T}d\tau\
e^{-i\Omega_n\tau}\
\nonumber\\
&& \hskip 60pt \times \chi_{\text{s}}^{ij}({\bm x},{\bm y};\tau),
\label{eq:4.41}
\eea
we find that $\chi_{\text{s}}^{+-}$ and $\chi_{\text{s}}^{-+}$ reflect the
strongest hydrodynamic contribution, which is given in terms of the
$\phi$-$\phi$ correlation function at wave vector $\pm{\bm q}$,
\bse
\label{eqs:4.42}
\bea
\chi_{\text{s}}^{\pm\mp}({\bm k},{\bm p};i\Omega_n) &=& \delta_{{\bm k}{\bm
p}}\,m_0^2\,\chi_{\phi\phi}({\bm k}\pm{\bm q},i\Omega_n)
\nonumber\\
&& \hskip 0pt +\ ({\text{less leading terms}}).
\label{eq:4.42a}
\eea
The ``less leading terms'' in Eq.\ (\ref{eq:4.42a}) reflect the terms
proportional to $c$ in Eq.\ (\ref{eq:4.38a}). They are either less singular
than the leading term, which scales as $1/\omega^2$, or have a prefactor that
is small by a factor of $cq\propto q^2$. $\chi_{\text{s}}^{++}$ and
$\chi_{\text{s}}^{--}$ also are proportional to $\chi_{\phi\phi}$, but they are
not diagonal in the momenta,
\be
\chi_{\text{s}}^{\pm\pm}({\bm k},{\bm p};i\Omega_n) = - \delta_{{\bm p},{\bm
k}\pm 2{\bm q}}\,m_0^2\,\chi_{\phi\phi}({\bm k}\pm{\bm q},i\Omega_n).
\label{eq:4.42b}
\ee
\ese
Only the terms with ${\bm k}={\bm p}$ contribute to the neutron scattering
cross section. Right and left circularly polarized neutrons will therefore see
the hydrodynamic singularity only at wave vector ${\bm k}={\bm q}$ and ${\bm
k}=-{\bm q}$, respectively. Unpolarized neutrons will see symmetric
contributions at ${\bm k} = \pm{\bm q}$. For instance, the $xx$ component of
the susceptibility tensor, $\chi_{\text{s}}^{11} = (\chi_{\text{s}}^{+-} +
\chi_{\text{s}}^{-+} + \chi_{\text{s}}^{++} + \chi_{\text{s}}^{--})/4$ is given
by
\begin{widetext}
\be
\chi_{\text{s}}^{11}({\bm k},{\bm p};i\Omega_n) =
\frac{m_0^2}{4}\,\Bigl[\delta_{{\bm p}{\bm k}}\bigl(\chi_{\phi\phi}({\bm
k}+{\bm q},i\Omega_n) + \chi_{\phi\phi}({\bm k}-{\bm q},i\Omega_n)\bigr) -
\delta_{{\bm p},{\bm k}+2{\bm q}}\,\chi_{\phi\phi}({\bm k}+{\bm q},i\Omega_n) -
\delta_{{\bm p},{\bm k}-2{\bm q}}\,\chi_{\phi\phi}({\bm k}-{\bm
q},i\Omega_n)\Bigr],
\label{eq:4.43}
\ee
\end{widetext}
with the first term in the square brackets contributing to the neutron
scattering cross section.

The longitudinal component $\chi_{\text{s}}^{33}$ is less singular than
$\chi_{\text{s}}^{+-}$ since the contribution of $\phi$ to $(A^{-1}\delta M)_z$
is suppressed by a transverse gradient, see Eqs.\ (\ref{eq:4.38b},
\ref{eq:4.39a}). Accordingly, the leading hydrodynamic contribution to
$\chi_{\text{s}}^{33}$ at ${\bm k} = \pm {\bm q}$ scales as
$\langle\varphi_1\varphi_1\rangle \sim 1/\omega$. There is also a weak
signature of the Goldstone mode in the vicinity of ${\bm k} = 0$ due to the
contribution of $\langle\pi_1\pi_1\rangle$ to $\chi_{\text{s}}^{33}$, but this
scales only as $\omega^0$.

All of these results are in agreement with what one expects from Sec.\
\ref{subsec:II.D}.

\section{Discussion and conclusion}
\label{sec:V}

Let us summarize our results. We have developed a framework for a theoretical
description of itinerant quantum helimagnets that is analogous to Hertz's
treatment of ferromagnets.\cite{Hertz_1976} We have analyzed this theory in the
helically ordered phase at a mean-field/Gaussian level analogous to Stoner
theory. As in the ferromagnetic case there is a split Fermi surface, with the
splitting proportional to the amplitude of the helically modulated
magnetization. We then focused on the Goldstone mode, or helimagnon, that
results from the spontaneous breaking of the translational symmetry in the
helical phase. We have found that the helimagnon is a propagating, weakly
damped mode with a strongly anisotropic dispersion relation. The frequency
scales linearly with the wave number for wave vectors parallel to the pitch of
the helix, but quadratically for wave vectors in the transverse direction. In
this sense the helimagnon behaves like an antiferromagnetic magnon in the
longitudinal direction, but like a ferromagnetic one in the transverse
direction. This anisotropy is analogous to the situation in chiral liquid
crystals, and indeed the results of our microscopic theory are qualitatively
reproduced by combining an educated guess of the statics, inferred from the
liquid-crystal case, with a phenomenological time-dependent Ginzburg-Landau
theory for the dynamics. The structure of the microscopic field theory is in
one-to-one correspondence with the structure of the phenomenological theory.
Particle-hole excitations provide a damping of the helimagnon, with a damping
coefficient proportional to the pitch wave number to the fourth power.

Our continuum theory ignores the spin-orbit coupling of the electron spins to
the underlying lattice which, in conjunction with crystal-field effects, will
change the dispersion relation of the Goldstone mode at very small wave numbers
or frequencies. In contrast to the case of ferromagnetic and antiferromagnetic
magnons, however, breaking the spin rotational symmetry does not give the
helimagnons a mass; it just changes the dispersion relation at unobservably
small wave numbers, making it less soft. Consistent with this, soft helimagnons
are observable by neutron scattering,\cite{Roessli_et_al_2004} as are
ferromagnetic and antiferromagnetic magnons. The neutron scattering cross
section is proportional to the magnetic structure factor, which in turn is
simply related to the spin susceptibility, which we have shown to be
proportional to the helimagnon correlation function.

Let us now discuss some observable properties, using parameter values
appropriate for MnSi. MnSi has a Fermi temperature $T_{\text{F}} \approx
23,200\,\text{K}$,\cite{Nakanishi_Yanase_Hasegawa_1980} a pitch wave number $q
\approx 0.035\, \text{\AA}^{-1}$,\cite{Ishikawa_et_al_1985} and an effective
electron mass, averaged over the Fermi surface, $m_{\text{e}} \approx 4
m_0$,\cite{Taillefer_Lonzarich_Strange_1986} with $m_0$ the free electron mass.
In a nearly-free-electron model this leads to $k_{\text{F}} \approx
1.45\,\text{\AA}^{-1}$ and $qv_{\text{F}}/k_{\text{B}} \approx
1,000\,\text{K}$. The value of the exchange splitting $\lambda$ is less clear.
The large ordered moment of about $0.4\,\mu_{\text{B}}$ per formula
unit\cite{Pfleiderer_et_al_2004} suggests an exchange splitting that is a
substantial fraction of $T_{\text{F}}$. This is hard to reconcile with the low
Curie temperature $T_{\text{C}} = 29.5\,{\text{K}}$ at ambient pressure. Ref.\
\onlinecite{Taillefer_Lonzarich_Strange_1986} found an exchange splitting
$\lambda/k_{\text{B}} \approx 520\,\text{K}$, which is hard to reconcile with
the large ordered moment. We recall that the helical order is caused by the
(weak) spin-orbit interaction, which suggests a clear separation of energy
scales, and in particular $\lambda \gg qv_{\text{F}}$. In judging such
estimates one needs to keep in mind that MnSi is a fairly strong magnet (as
evidenced by the large ordered moment) with a complicated Fermi surface with
both electron and hole orbits, and a nearly-free-electron model as well as a
weak-coupling Stoner picture are of limited applicability.

One question that arises is the value of $qv_{\text{F}}/\lambda$ appropriate
for MnSi (recall that our calculation is for $qv_{\text{F}}/\lambda \ll 1$).
This hinges on the value of $\lambda$. If one accepts a sizeable value of
$\lambda$ (in units of the Fermi temperature) as suggested by the large value
of the ordered magnetic moment, then $qv_{\text{F}}/\lambda \ll 1$. If one
accepts the much smaller value for $\lambda$ obtained in Ref.\
\onlinecite{Taillefer_Lonzarich_Strange_1986}, then $qv_{\text{F}}/\lambda
\approx 1$. Even in the latter case we expect our considerations to still apply
qualitatively, although the lack of a clear separation of energy scales would
make a quantitative analysis more difficult.

The anisotropic helimagnon is well defined for wave numbers $\vert{\bm k}\vert
< q$. In MnSi, $q \approx 0.035\,\text{\AA}^{-1}$, which is accessible by
inelastic neutron scattering.\cite{Pfleiderer_et_al_2004} An estimate of the
helimagnon excitation energy in MnSi at this wave number can be obtained from
Ref.\ \onlinecite{Semadeni_et_al_1999}. In this experiment, a small magnetic
field was used to destroy the helix. The resulting ferromagnetic magnons at a
wave number equal to $q$ had an energy of about 300 mK above the field-induced
gap. An estimate of the helimagnon energy at the same wave number from Eq.\
(\ref{eq:4.33b}) yields the same order of magnitude if $\lambda$ is a
substantial fraction of the Fermi energy. The prediction is thus for an
anisotropic helimagnon, which at a wave number on the order of $q$, and an
energy on the order of 300 mK, will cross over to a ferromagnetic magnon. The
damping is expected to be weak, especially in systems with some quenched
disorder, although in ultraclean systems it may be strongly wave vector
dependent.\cite{damping_footnote}

One expects the low-energy helimagnon mode to have an appreciable effect on
other observables such as the specific heat, or the electrical resistivity.
This is indeed the case, and we will discuss these effects in a separate
paper.\cite{paper_II}

In conclusion, the present paper provides a general many-body formalism for
itinerant quantum helimagnets, which can be used for calculating any observable
of interest, with the single-particle Green function and the helimagnon
propagator as building blocks. Observables of obvious interest include the
specific heat, the quasi-particle relaxation time, and the resistivity.
Calculations of these quantities within the framework of the present theory
will be reported in a separate paper.\cite{paper_II}

\acknowledgments We thank the participants of the Workshop on Quantum Phase
Transitions at the KITP at UCSB, and in particular Christian Pfleiderer and
Thomas Vojta, for stimulating discussions. DB would like to thank Peter
B{\"o}ni, Christian Pfleiderer, and Willi Zwerger for their hospitality during
a visit to the Technical University Munich. This work was supported by the NSF
under grant Nos. DMR-01-32555, DMR-01-32726, and PHY-99-07949, and by the DFG
under grant No. SFB 608.

\appendix

\section{Rotational-invariance-breaking terms in the Ginzburg Landau expansion}
\label{app:X}

In addition to the term considered in Eq.\ (\ref{eq:2.21}), there are other
terms that break the rotational symmetry and contribute to the same order in
the spin-orbit interaction strength $g_{\text{\,SO}}$. The detailed structure
of such terms in the action depends on the precise lattice structure. For
definiteness we assume the cubic P2$_1$3 space group realized, for instance, in
MnSi.

The leading terms which break the rotational symmetry are of the form
\cite{Bak_Jensen_1980, Nakanishi_et_al_1980}
\begin{widetext}
\bea
\delta S &=& \int d{\bm x}\ \Bigl\{a_1 \left[( \partial_x^2 {\bm M}({\bm
x}))^2+(
\partial_y^2 {\bm M}({\bm
x}))^2+ (
\partial_z^2 {\bm M}({\bm
x}))^2 \right] + g_{\text{\,SO}}^2 \left[a_2 (\partial_x M_y({\bm x}))^2 + a_3
(\partial_x M_z({\bm x}))^2 + {\text{cycl.}}\right]
\nonumber\\
&& \hskip 50pt +\ g_{\text{\,SO}}^4 a_4 \left[ (M_x({\bm x}))^4 + (M_y({\bm
x}))^4 + (M_z({\bm x}))^4\right]\Bigr\}.
\label{eq:X.1}
\eea
\end{widetext}
Here $a_1$, $a_2$, $a_3$ and $a_4$ are constants which remain finite for
vanishing spin-orbit coupling, $g_{\text{\,SO}} \to 0$, and cycl. denotes
cyclic permutations of $x$, $y$, and $z$. Omitted terms like $(M_x^2 M_y^2 +
{\text{cycl.}})$ or $[(\partial_x M_x)^2 + {\text{cycl.}}]$ can be obtained by
adding rotationally invariant terms to $\delta S$. We note that the
Dzyaloshinski-Moriya interaction, i.e. the constant $c$ in Eq.~(\ref{eq:2.1}),
the pitch wave number $q$, and therefore all typical momenta, are {\em linear}
in $g_{\text{\,SO}}$. Consequently, all terms in Eq.\ (\ref{eq:X.1}) contribute
only to order $g_{\text{\,SO}}^4$. They are therefore small compared to the
energy gain obtained by forming the helical state, which is is of order $q^2
\propto g_{\text{\,SO}}^2$.

The main effect of $\delta S$  is to pin the direction of the spiral to some
high-symmetry direction, namely, either (1,1,1) and equivalent directions, or
(1,0,0) and equivalent directions.\cite{Bak_Jensen_1980, Nakanishi_et_al_1980}
In addition, they change the dispersion relation of the helical Goldstone mode
at extremely small wave numbers. For one particular term this has been
demonstrated in Sec.\ \ref{subsec:II.E}; all other terms have qualitatively the
same effect. Notice that in a {\em ferro}magnet the effect on the Goldstone
mode is stronger. For instance, the cubic anisotropy (the term with coupling
constant $a_4$ in Eq.\ (\ref{eq:X.1})) gives the ferromagnetic magnons a true
mass, leaving no soft modes. This is a result of the fact that in a
ferromagnetic state the translational invariance is not spontaneously broken,
while in a helimagnetic one it is.

\begin{widetext}
\section{The reference ensemble spin susceptibility}
\label{app:A}

Substituting Eq.\ (\ref{eq:4.13a}) in Eq.\ (\ref{eq:4.14b}) and performing the
spin traces yields
\bea
\chi_0^{ij}({\bm k},{\bm p};i\Omega_n) &=& \frac{-1}{V}\sum_{{\bm k}'}
 T\sum_{i\omega_m} \biggl\{ \delta_{{\bm k},{\bm p}}\Bigl[
\bigl(\Sigma_{aa}^{++}\bigr)_{ij}\,a_+({\bm k}'-{\bm k},{\bm q};i\omega_m -
i\Omega_n)\,a_+({\bm k}',{\bm q};i\omega_n)
\nonumber\\
&& \hskip 78pt + \bigl(\Sigma_{aa}^{+-}\bigr)_{ij}\,a_+({\bm k}'-{\bm k},{\bm
q};i\omega_m - i\Omega_n)\,a_-({\bm k}',{\bm q};i\omega_n)
\nonumber\\
&& \hskip 78pt + \bigl(\Sigma_{aa}^{-+}\bigr)_{ij}\,a_-({\bm k}'-{\bm k},{\bm
q};i\omega_m - i\Omega_n)\,a_+({\bm k}',{\bm q};i\omega_n)
\nonumber\\
&& \hskip 78pt + \bigl(\Sigma_{aa}^{--}\bigr)_{ij}\,a_-({\bm k}'-{\bm k},{\bm
q};i\omega_m - i\Omega_n)\,a_-({\bm k}',{\bm q};i\omega_n)
\nonumber\\
&& \hskip 78pt + \bigl(\Sigma_{bb}^{+-}\bigr)_{ij}\,b_+({\bm k}'-{\bm k},{\bm
q};i\omega_m - i\Omega_n)\,b_-({\bm k}'+{\bm q},{\bm q};i\omega_n)
\nonumber\\
&& \hskip 78pt + \bigl(\Sigma_{bb}^{-+}\bigr)_{ij}\,b_-({\bm k}'-{\bm k},{\bm
q};i\omega_m - i\Omega_n)\,b_+({\bm k}'-{\bm q},{\bm q};i\omega_n) \Bigr]
\nonumber\\
&& \hskip 46pt + \delta_{{\bm k}-{\bm q},{\bm p}}\Bigl[
\bigl(\Sigma_{ab}^{++}\bigr)_{ij}\,a_+({\bm k}'-{\bm k},{\bm q};i\omega_m -
i\Omega_n)\,b_+({\bm k}'-{\bm q},{\bm q};i\omega_n)
\nonumber\\
&& \hskip 78pt + \bigl(\Sigma_{ab}^{-+}\bigr)_{ij}\,a_-({\bm k}'-{\bm k},{\bm
q};i\omega_m - i\Omega_n)\,b_+({\bm k}'-{\bm q},{\bm q};i\omega_n)
\nonumber\\
&& \hskip 78pt + \bigl(\Sigma_{ba}^{++}\bigr)_{ij}\,b_+({\bm k}'-{\bm k},{\bm
q};i\omega_m - i\Omega_n)\,a_+({\bm k}',{\bm q};i\omega_n)
\nonumber\\
&& \hskip 78pt + \bigl(\Sigma_{ba}^{+-}\bigr)_{ij}\,b_+({\bm k}'-{\bm k},{\bm
q};i\omega_m - i\Omega_n)\,a_-({\bm k}',{\bm q};i\omega_n) \Bigr]
\nonumber\\
&& \hskip 46pt + \delta_{{\bm k}+{\bm q},{\bm p}}\Bigl[
\bigl(\Sigma_{ab}^{+-}\bigr)_{ij}\,a_+({\bm k}'-{\bm k},{\bm q};i\omega_m -
i\Omega_n)\,b_-({\bm k}'+{\bm q},{\bm q};i\omega_n)
\nonumber\\
&& \hskip 78pt + \bigl(\Sigma_{ab}^{--}\bigr)_{ij}\,a_-({\bm k}'-{\bm k},{\bm
q};i\omega_m - i\Omega_n)\,b_-({\bm k}'+{\bm q},{\bm q};i\omega_n)
\nonumber\\
&& \hskip 78pt + \bigl(\Sigma_{ba}^{-+}\bigr)_{ij}\,b_-({\bm k}'-{\bm k},{\bm
q};i\omega_m - i\Omega_n)\,a_+({\bm k}',{\bm q};i\omega_n)
\nonumber\\
&& \hskip 78pt + \bigl(\Sigma_{ba}^{--}\bigr)_{ij}\,b_-({\bm k}'-{\bm k},{\bm
q};i\omega_m - i\Omega_n)\,a_-({\bm k}',{\bm q};i\omega_n) \Bigr]
\nonumber\\
&& \hskip 46pt + \delta_{{\bm k}-2{\bm q},{\bm p}}\
\bigl(\Sigma_{bb}^{++}\bigr)_{ij}\,b_+({\bm k}'-{\bm k},{\bm q};i\omega_m -
i\Omega_n)\,b_+({\bm k}'-{\bm q},{\bm q};i\omega_n)
\nonumber\\
&& \hskip 46pt + \delta_{{\bm k}+2{\bm q},{\bm p}}\
\bigl(\Sigma_{bb}^{--}\bigr)_{ij}\,b_-({\bm k}'-{\bm k},{\bm q};i\omega_m -
i\Omega_n)\,b_-({\bm k}'+{\bm q},{\bm q};i\omega_n)\ \biggr\}.
\nonumber\\
\label{eq:A.1}
\eea
\end{widetext}
The $\Sigma$ symbols denote traces of Pauli matrices
\bse
\label{eqs:A.2}
\be
\bigl(\Sigma_{aa}^{++}\bigr)_{ij} = \tr \bigl(
\sigma_i\,\sigma_+\,\sigma_-\,\sigma_j\,\sigma_+\sigma_- \bigr) =
   \left(\begin{array}{ccc} 0 & 0 & 0 \\
                            0 & 0 & 0 \\
                            0 & 0 & 1 \end{array}\right),
\label{eq:A.2a}
\ee
\be
\bigl(\Sigma_{aa}^{+-}\bigr)_{ij} = \tr \bigl(
\sigma_i\,\sigma_+\,\sigma_-\,\sigma_j\,\sigma_-\sigma_+ \bigr) =
   \left(\begin{array}{ccc} 1 & -i & 0 \\
                            i &  1 & 0 \\
                            0 & 0  & 0    \end{array}\right),
\label{eq:A.2b}
\ee
\be
\bigl(\Sigma_{aa}^{-+}\bigr)_{ij} = \tr \bigl(
\sigma_i\,\sigma_-\,\sigma_+\,\sigma_j\,\sigma_+\sigma_- \bigr) =
   \left(\begin{array}{ccc} 1  & i & 0 \\
                            -i & 1 & 0 \\
                            0  & 0 & 0    \end{array}\right),
\label{eq:A.2c}
\ee
\be
\bigl(\Sigma_{aa}^{--}\bigr)_{ij} = \tr \bigl(
\sigma_i\,\sigma_-\,\sigma_+\,\sigma_j\,\sigma_-\sigma_+ \bigr) =
   \left(\begin{array}{ccc} 0 & 0 & 0 \\
                      0 & 0 & 0 \\
                      0 & 0 & 1 \end{array}\right),
\label{eq:A.2d}
\ee
\be
\bigl(\Sigma_{bb}^{+-}\bigr)_{ij} = \tr \bigl(
\sigma_i\,\sigma_+\,\sigma_j\,\sigma_- \bigr) =
   \left(\begin{array}{ccc} 0 & 0 & 0 \\
                            0 & 0 & 0 \\
                            0 & 0 & -1  \end{array}\right),
\label{eq:A.2e}
\ee
\be
\bigl(\Sigma_{bb}^{-+}\bigr)_{ij} = \tr \bigl(
\sigma_i\,\sigma_-\,\sigma_j\,\sigma_+ \bigr) =
   \left(\begin{array}{ccc} 0 & 0 & 0 \\
                            0 & 0 & 0 \\
                            0 & 0 & -1  \end{array}\right),
\label{eq:A.2f}
\ee
\be
\bigl(\Sigma_{ab}^{++}\bigr)_{ij} = \tr \bigl(
\sigma_i\,\sigma_+\,\sigma_-\,\sigma_j\,\sigma_+ \bigr) =
   \left(\begin{array}{ccc} 0 & 0 & 1 \\
                            0 & 0 & i \\
                            0 & 0 & 0  \end{array}\right),
\label{eq:A.2g}
\ee
\be
\bigl(\Sigma_{ab}^{-+}\bigr)_{ij} = \tr \bigl(
\sigma_i\,\sigma_-\,\sigma_+\,\sigma_j\,\sigma_+ \bigr) =
   \left(\begin{array}{ccc} 0 & 0 & 0 \\
                            0 & 0 & 0 \\
                            -1 & -i & 0  \end{array}\right),
\label{eq:A.2h}
\ee
\be
\bigl(\Sigma_{ba}^{++}\bigr)_{ij} = \tr \bigl(
\sigma_i\,\sigma_+\,\sigma_j\,\sigma_+\,\sigma_- \bigr) =
   \left(\begin{array}{ccc} 0 & 0 & 0 \\
                            0 & 0 & 0 \\
                            1 & i & 0  \end{array}\right),
\label{eq:A.2i}
\ee
\be
\bigl(\Sigma_{ba}^{+-}\bigr)_{ij} = \tr \bigl(
\sigma_i\,\sigma_+\,\sigma_j\,\sigma_-\,\sigma_+ \bigr) =
   \left(\begin{array}{ccc} 0 & 0 & -1 \\
                            0 & 0 & -i \\
                            0 & 0 & 0  \end{array}\right),
\label{eq:A.2j}
\ee
\be
\bigl(\Sigma_{ab}^{+-}\bigr)_{ij} = \tr \bigl(
\sigma_i\,\sigma_+\,\sigma_-\,\sigma_j\,\sigma_- \bigr) =
   \left(\begin{array}{ccc} 0 & 0 & 0 \\
                            0 & 0 & 0 \\
                            1 & -i & 0  \end{array}\right),
\label{eq:A.2k}
\ee
\be
\bigl(\Sigma_{ab}^{--}\bigr)_{ij} = \tr \bigl(
\sigma_i\,\sigma_-\,\sigma_+\,\sigma_j\,\sigma_- \bigr) =
   \left(\begin{array}{ccc} 0 & 0 & -1 \\
                            0 & 0 & i \\
                            0 & 0 & 0  \end{array}\right),
\label{eq:A.2l}
\ee
\be
\bigl(\Sigma_{ba}^{-+}\bigr)_{ij} = \tr \bigl(
\sigma_i\,\sigma_-\,\sigma_j\,\sigma_+\,\sigma_- \bigr) =
   \left(\begin{array}{ccc} 0 & 0 & 1 \\
                            0 & 0 & -i \\
                            0 & 0 & 0  \end{array}\right),
\label{eq:A.2m}
\ee
\be
\bigl(\Sigma_{ba}^{--}\bigr)_{ij} = \tr \bigl(
\sigma_i\,\sigma_-\,\sigma_j\,\sigma_-\,\sigma_+ \bigr) =
   \left(\begin{array}{ccc} 0 & 0 & 0 \\
                            0 & 0 & 0 \\
                            -1 & i & 0  \end{array}\right),
\label{eq:A.2n}
\ee
\be
\bigl(\Sigma_{bb}^{++}\bigr)_{ij} = \tr \bigl(
\sigma_i\,\sigma_+\,\sigma_j\,\sigma_+ \bigr) =
   \left(\begin{array}{ccc} 1 & i & 0 \\
                            i & -1 & 0 \\
                            0 & 0 & 0  \end{array}\right),
\label{eq:A.2o}
\ee
\be
\bigl(\Sigma_{bb}^{--}\bigr)_{ij} = \tr \bigl(
\sigma_i\,\sigma_-\,\sigma_j\,\sigma_- \bigr) =
   \left(\begin{array}{ccc} 1 & -i & 0 \\
                            -i & -1 & 0 \\
                            0 & 0 & 0  \end{array}\right),
\label{eq:A.2p}
\ee
\ese

\section{The functions $f_{\phi\phi}$, $f_{11}$, $f_{12}$, $g_{11}$ and $h_{\phi 1}$}
\label{app:B}

In this appendix we show how to evaluate the functions $f_{\phi\phi}$,
$f_{11}$, $f_{12}$, and $h_{\phi 1}$ defined in Eqs.\ (\ref{eq:4.18c} -
\ref{eq:4.18g}) and (\ref{eqs:4.30}) in the limit of long wavelengths and small
frequencies.

\subsection{The function $f_{\phi\phi}$}
\label{subapp:B.1}

We start with the expression for $\varphi_{\phi\phi}$ in terms of Green
functions given in Eq.\ (\ref{eq:4.18f}). By symmetrizing the dependence on
${\bm k}$ and ${\bm q}$, performing the sum over Matsubara frequencies, and
doing a partial fraction decomposition, $\varphi_{\phi\phi}$ can be written
\bse
\label{eqs:B.1}
\bea
\varphi_{\phi\phi}({\bm k},i\Omega_n) &=& \varphi_{\phi\phi}^{(+)}({\bm
k},i\Omega_n) + \varphi_{\phi\phi}^{(+)}(-{\bm k},-i\Omega_n)
\nonumber\\
&& + \varphi_{\phi\phi}^{(-)}({\bm k},i\Omega_n) -
\varphi_{\phi\phi}^{(-)}(-{\bm k},-i\Omega_n).
\nonumber\\
\label{eq:B.1a}
\eea
From Eq.\ (\ref{eq:4.18c}) we have
\be
f_{\phi\phi}({\bm k},i\Omega_n) = 2\left[\varphi_{\phi\phi}^{(+)}({\bm
k},i\Omega_n) + \varphi_{\phi\phi}^{(+)}(-{\bm k},-i\Omega_n)\right].
\label{eq:B.1b}
\ee
That is, $f_{\phi\phi}$ is given by the symmetric part of $\varphi_{\phi\phi}$
alone. The antisymmetric part $\varphi_{\phi\phi}^{(-)}$ does not contribute,
but we list it here for completeness:
\begin{widetext}
\bea
\varphi_{\phi\phi}^{(-)}({\bm k},i\Omega_n) &=& \frac{-1}{4V}\sum_{\bm
p}\biggl\{\frac{({\bm q}\cdot{\bm p}_-)w_+ - ({\bm q}\cdot{\bm
p}_+)w_-}{2m_{\text{e}}w_+w_-}\biggl[\frac{f(\xi^q_- - w_-)}{i\Omega_n - {\bm
k}\cdot{\bm p}/m_{\text{e}} + w_+ - w_-} - \frac{f(\xi^q_- + w_-)}{i\Omega_n -
{\bm k}\cdot{\bm p}/m_{\text{e}} - w_+ + w_-}\biggr]
\nonumber\\
&& + \frac{({\bm q}\cdot{\bm p}_-)w_+ + ({\bm q}\cdot{\bm
p}_+)w_-}{2m_{\text{e}}w_+w_-}\biggl[\frac{f(\xi^q_- - w_-)}{i\Omega_n - {\bm
k}\cdot{\bm p}/m_{\text{e}} - w_+ - w_-} - \frac{f(\xi^q_- + w_-)}{i\Omega_n -
{\bm k}\cdot{\bm p}/m_{\text{e}} + w_+ + w_-}\biggr]\biggr\}.
\label{eq:B.1c}
\eea
The symmetric part, $\varphi^{(+)}$, consists of two parts that are
structurally distinct,
\be
\varphi_{\phi\phi}^{(+)}({\bm k},i\Omega_n) = \varphi_{\phi\phi}^{(1)}({\bm
k},i\Omega_n) + \varphi_{\phi\phi}^{(2)}({\bm k},i\Omega_n),
\label{eq:B.1d}
\ee
where
\be
\varphi_{\phi\phi}^{(1)}({\bm k},i\Omega_n) = \frac{-1}{4V}\sum_{\bm p}
\frac{w_+w_- + \lambda^2 + g({\bm p})}{w_+w_-}\biggl[\frac{f(\xi^q_- -
w_-)}{i\Omega_n - {\bm k}\cdot{\bm p}/m_{\text{e}} - w_+ - w_-} +
\frac{f(\xi^q_- + w_-)}{i\Omega_n - {\bm k}\cdot{\bm p}/m_{\text{e}} + w_+
+w_-}\biggr],
\label{eq:B.1e}
\ee
\be
\varphi_{\phi\phi}^{(2)}({\bm k},i\Omega_n) = \frac{-1}{4V}\sum_{\bm
p}\frac{w_+w_- - \lambda^2 - g({\bm p})}{w_+w_-}\biggl[\frac{f(\xi^q_- -
w_-)}{i\Omega_n - {\bm k}\cdot{\bm p}/m_{\text{e}} + w_+ - w_-} +
\frac{f(\xi^q_- + w_-)}{i\Omega_n - {\bm k}\cdot{\bm p}/m_{\text{e}} - w_+
+w_-}\biggr].
\label{eq:B.1f}
\ee
\end{widetext}
In these expressions $\xi^q_{\pm} = \xi_{{\bm p}_{\pm}} + {\bm
q}^2/8m_{\text{e}}$, ${\bm p}_{\pm} = {\bm p} \pm {\bm k}/2$, and $w_{\pm} =
w({\bm p}_{\pm})$ with
\be
w({\bm p}) = \sqrt{\frac{({\bm q}\cdot{\bm p})^2}{4m_{\text{e}}^2} +
\lambda^2},
\label{eq:B.1g}
\ee
and
\be
g({\bm p}) = \frac{1}{4m_{\text{e}}^2}\left(({\bm q}\cdot{\bm p})^2 -
\frac{1}{4}\,({\bm q}\cdot{\bm k})^2\right).
\label{eq:B.1h}
\ee
\ese

$\varphi_{\phi\phi}^{(1)}$ and $\varphi_{\phi\phi}^{(2)}$ are both generalized
Lindhard functions. For $\lambda=0$ they combine to form a Lindhard function at
wave vector ${\bm k} + {\bm q}$, while for ${\bm q}=0$
$\varphi_{\phi\phi}^{(2)}$ vanishes and $\varphi_{\phi\phi}^{(1)}$ turns into
the function that determines the ferromagnetic magnon\cite{Belitz_et_al_1998}.
A structural difference between them is that in $\varphi_{\phi\phi}^{(1)}$ the
hydrodynamic singularity at ${\bm k} = i\Omega_n = 0$ that is characteristic
for the Lindhard function is protected by $\lambda$, while in
$\varphi_{\phi\phi}^{(2)}$ this is not the case.

The remaining wave vector integral is difficult, and we evaluate it only in the
limit $\lambda \gg qv_{\text{F}}$. The Fermi functions in Eqs.\
(\ref{eq:B.1e},\ref{eq:B.1f}) pin the integration momentum ${\bm p}$ to a
shifted Fermi surface. For $\lambda > qv_{\text{F}}$ one can therefore perform
a straightforward expansion of the integrand in powers of
$qv_{\text{F}}/\lambda$. For $\varphi^{(1)}$ the leading term is the ${\bm
q}=0$ contribution. Specifically, in this limit
\be
(w_+ w_- + \lambda^2 + g({\bm p}))/w_+ w_- \to 2,
\label{eq:B.2}
\ee
and the calculation reduces to the ferromagnetic case. For $k \ll k_{\text{F}}$
and $\vert\Omega_n\vert \ll \lambda$, the result is
\be
\varphi_{\phi\phi}^{(1)}({\bm k},i\Omega_n) = \varphi_{\phi\phi}^{(1)}(0,i0) +
N_{\text{F}}\left[-\frac{1}{3}\,\left(\frac{{\bm k}}{2k_{\text{F}}}\right)^2 +
\left(\frac{i\Omega_n}{2\lambda}\right)^2\right].
\label{eq:B.3}
\ee

For $\varphi^{(2)}$ the leading term is of $O(q^2)$, since
\be
(w_+ w_- - \lambda^2 - g({\bm p}))/w_+ w_- \to ({\bm q}\cdot{\bm
k})^2/8m_{\text{e}}^2\lambda^2,
\label{eq:B.4}
\ee
and the integral is reduces to a Lindhard function. The non-hydrodynamic part
provides only corrections of $O((qv_{\text{F}}/\lambda)^2)$ to $\varphi^{(1)}$,
but the hydrodynamic part is qualitatively new and must be kept. We find
\be
\varphi^{(2)}({\bm k},i\Omega_n) = \varphi^{(2)}({\bm k},i0) -
N_{\text{F}}\,\frac{\pi}{32}\,\frac{({\bm q}\cdot{\bm
k})^2}{m_{\text{e}}^2\lambda^2}\,\frac{\vert\Omega_n\vert}{v_{\text{F}}\vert{\bm
k}\vert}\ ,
\label{eq:B.5}
\ee
which is valid for $\vert\Omega_n\vert \ll v_{\text{F}}k$. Combining these
result yields Eq.\ (\ref{eq:4.20a}) with coefficients appropriate for the case
$\lambda >> qv_{\text{F}}$. Corrections are of $O((qv_{\text{F}}/\lambda)^2)$.

\subsection{The functions $f_{11}$, $f_{12}$, and $g_{11}$}
\label{subapp:B.2}

In order to calculate $\varphi_{11}$, Eq.\ (\ref{eq:4.18g}), it is convenient
to consider consider $\varphi_{11}({\bm k}-{\bm q},i\Omega_n)$, which can be
written in a form similar to $\varphi_{\phi\phi}({\bm k},i\Omega_n)$, namely,
\begin{widetext}
\bse
\label{eqs:B.6}
\be
\varphi_{11}({\bm k}-{\bm q},i\Omega_n) = \varphi_{11}^{(1)}({\bm k}-{\bm
q},i\Omega_n) + \varphi_{11}^{(2)}({\bm k}-{\bm q},i\Omega_n),
\label{eq:B.6a}
\ee
where
\be
\varphi_{11}^{(1)}({\bm k}-{\bm q},i\Omega_n) = \frac{-1}{8V}\sum_{\bm p}
\frac{w_+ w_- - g({\bm p}) + \lambda^2}{w_+ w_-}\,\biggl[ \frac{f(\xi^q_- -
w_-) - f(\xi^q_+ + w_+)}{i\Omega_n - {\bm k}\cdot{\bm p}/m_{\text{e}} - w_+ -
w_-} + \frac{f(\xi^q_- + w_-) - f(\xi^q_+ - w_+)}{i\Omega_n - {\bm k}\cdot{\bm
p}/m_{\text{e}} + w_+ + w_-}\biggr]\ ,
\label{eq:B.6b}
\ee
\be
\varphi_{11}^{(2)}({\bm k}-{\bm q},i\Omega_n) = \frac{-1}{8V}\sum_{\bm p}
\frac{w_+ w_- + g({\bm p}) - \lambda^2}{w_+ w_-}\,\biggl[ \frac{f(\xi^q_- -
w_-) - f(\xi^q_+ - w_+)}{i\Omega_n - {\bm k}\cdot{\bm p}/m_{\text{e}} + w_+ -
w_-} + \frac{f(\xi^q_- + w_-) - f(\xi^q_+ + w_+)}{i\Omega_n - {\bm k}\cdot{\bm
p}/m_{\text{e}} - w_+ + w_-}\biggr]\ .
\label{eq:B.6c}
\ee
\ese
\end{widetext}
To evaluate these integrals we again assume $\lambda \gg qv_{\text{F}}$. In
this limit
\be
(w_+ w_- - g({\bm p}) + \lambda^2)/w_+ w_- \to 2,
\label{eq:B.7}
\ee
and the integral again reduces to the ferromagnetic case. The result is
\bse
\label{eqs:B.8}
\bea
\varphi_{11}^{(1)}({\bm k}-{\bm q},i\Omega_n) &=& \varphi_{11}({\bm q},i0)
\nonumber\\
&& \hskip -50pt  + \frac{N_{\text{F}}}{2}\,\Bigl[-\frac{1}{3}\,\left(\frac{{\bm
k}}{2k_{\text{F}}}\right)^2 + \left(\frac{i\Omega_n}{2\lambda}\right)^2\Bigr].
\label{eq:B.8a}
\eea
Shifting ${\bm k}$ by ${\bm q}$, we find
\bea
\varphi_{11}^{(1)}({\bm k},i\Omega_n) &=& \varphi_{11}^{(1)}(0,i0) \hskip 130pt
\nonumber\\
&& \hskip -60pt + \frac{N_{\text{F}}}{2}\,\left[-\,\frac{2}{3}\,\frac{{\bm
k}\cdot{\bm q}}{(2k_{\text{F}})^2} -
\frac{1}{3}\,\left(\frac{k}{2k_{\text{F}}}\right)^2 +
\left(\frac{i\Omega_n}{2\lambda}\right)^2\right]\ .
\label{eq:B.8b}
\eea
\ese
For $\varphi_{11}^{(2)}$ we need
\be
(w_+ w_- + g({\bm p}) - \lambda^2)/w_+ w_- \to ({\bm q}\cdot{\bm
p})^2/2m_{\text{e}}^2\lambda^2.
\label{eq:B.9}
\ee
In contrast to Eq.\ (\ref{eq:B.4}), this is quadratic in the integration
variable ${\bm p}$. $\varphi_{11}^{(2)}$ therefore is a stress correlation
function. It has a hydrodynamic contribution of the same functional form as the
Lindhard function, but only the transverse (with respect to ${\bm k}$)
components contribute to it. The non-hydrodynamic contributions are again
subleading compared to $\varphi_{11}^{(1)}$. One finds
\bse
\label{eqs:B.10}
\bea
\varphi_{11}^{(2)}({\bm k}-{\bm q},i\Omega_n) &=& \varphi_{11}^{(2)}({\bm
k}-{\bm q},i0)
\nonumber\\
&& \hskip -40pt + N_{\text{F}}\,\frac{\pi}{32}\,\frac{v_{\text{F}}q}{\lambda}\,
\frac{q{\bm k}_{\perp}^2}{\vert{\bm
k}\vert^3}\,\frac{\vert\Omega_n\vert}{\lambda},
\label{eq:B.10a}
\eea
and after shifting the momentum we obtain
\be
\varphi_{11}^{(2)}({\bm k},i\Omega_n) = \varphi_{11}^{(2)}({\bm k},i0) +
N_{\text{F}}\,\frac{\pi}{32}\,\frac{v_{\text{F}}q}{\lambda}\,\frac{{\bm
k}_{\perp}^2}{q^2}\,\frac{\vert\Omega_n\vert}{\lambda}\ .
\label{eq:B.10b}
\ee
\ese
Using these results in Eqs.\ (\ref{eq:4.18d}, \ref{eq:4.18e}) and
(\ref{eq:4.29c}) we obtain Eqs.\ (\ref{eq:4.20b}, \ref{eq:4.20c}) and
(\ref{eq:4.30a}), respectively.

\subsection{The functions $h_{\phi 1}$}
\label{subapp:B.3}

Finally, the function $\eta_{\phi 1}$ which is defined by Eq.\ (\ref{eq:4.29e})
and determines $h_{\phi 1}$ according to Eq.\ (\ref{eq:4.29d}), can be written
\bse
\label{eqs:B.11}
\be
\eta_{\phi 1}({\bm k},i\Omega_n) = \eta^{(1)}_{\phi 1}({\bm k},i\Omega_n) +
\eta^{(2)}_{\phi 1}({\bm k},i\Omega_n),
\label{eq:B.11a}
\ee
where
\begin{widetext}
\bea
\eta^{(1)}_{\phi 1}({\bm k},i\Omega_n) &=& \frac{-\lambda}{4V}\sum_{\bm p}
\frac{w_+ + w_-}{w_+ w_-}\,\left[ \frac{f(\xi^q_- - w_-) - f(\xi^q_+ +
w_+)}{i\Omega_n - {\bm p}\cdot{\bm k}/m_{\text{e}} - w_+ - w_-} -
\frac{f(\xi^q_- + w_-) - f(\xi^q_+ - w_+)}{i\Omega_n - {\bm p}\cdot{\bm
k}/m_{\text{e}} + w_+ + w_-}\right],
\label{eq:B.11b}\\
\eta^{(2)}_{\phi 1}({\bm k},i\Omega_n) &=& \frac{-\lambda}{4V}\sum_{\bm p}
\frac{w_+ - w_-}{w_+ w_-}\,\left[ \frac{f(\xi^q_- - w_-) - f(\xi^q_+ -
w_+)}{i\Omega_n - {\bm p}\cdot{\bm k}/m_{\text{e}} + w_+ - w_-} -
\frac{f(\xi^q_- + w_-) - f(\xi^q_+ + w_+)}{i\Omega_n - {\bm p}\cdot{\bm
k}/m_{\text{e}} + w_+ - w_-}\right],
\label{eq:B.11c}
\eea
\end{widetext}
Using the same techniques as above we find, in the limit $\lambda \gg
qv_{\text{F}}$,
\be
\eta^{(1)}_{\phi 1}({\bm k},i\Omega_n) =
-N_{\text{F}}\,\frac{i\Omega_n}{2\lambda}.
\label{eq:B.11d}
\ee
\ese
$\eta^{(2)}_{\phi 1}$ is proportional to $N_{\text{F}}(i\Omega_n/\lambda)({\bm
q}\cdot{\bm k})^2/k_{\text{F}}^2 {\bm k}^2$, and hence is small compared to
$\eta^{(1)}_{\phi 1}$ by a factor of $q^2/k_{\text{F}}^2$. These results yield
Eq.\ (\ref{eq:4.30b}).

%\bibliography{helimagnons}

\begin{thebibliography}{43}
\bibitem[{dyn()}]{dynamics_footnote}
\bibinfo{note}{This expectation pertains in particular to the static
  properties. The dynamics in general, and the damping of the mode in
  particular, can be different and will be affected by the presence of
  fermionic quasi-particles in metallic helimagnets.}
\bibitem[{pha()}]{phase_footnote}
\bibinfo{note}{For simplicity we ignore the difference between the phase $\phi$
  and the generalized phase $u$. This is of no consequence since the correct
  rotational invariance is implicit in the effective action
  $S_{\text{eff}}[u]$, which ultimately determines the dynamics of $u$.}
\bibitem[{mas()}]{mass_footnote}
\bibinfo{note}{In principle one can imagine $r_0$ to become negative as a
  function of the coupling strength, which could lead to co-existing
  ferromagnetic and helical order. We will see later that this does not happen
  in the specific model we will consider, and we ignore this possibility.}
\bibitem[{num()}]{number_soft_modes_footnote}
\bibinfo{note}{Note that restoring the rotational invariance by putting $b=0$
  does not lead to additional Goldstone modes. The reason is that small
  fluctuations $\delta\,{\bm q}$ of the pitch vector lead to large changes of
  the order parameter, $\delta\,{\bm M} \propto {\bm x}\,\delta\,{\bm q}$ for
  large ${\bm x}$. This is the same effect that ensures the absence of a ${\bm
  k}_{\perp}^2$ term in the dispersion relation, and it is reminiscient of the
  Anderson-Higgs mechanism.}
\bibitem[{var()}]{varphi_footnote}
\bibinfo{note}{This restriction is necessary to avoid an illegal increase in
  the number of degrees of freedom when the real field $\varphi$ is represented
  in terms of two real fields $\varphi_1$ and $\varphi_2$.}
\bibitem[{fm_()}]{fm_limit_footnote}
\bibinfo{note}{Here we keep only terms that couple flucuations at the same wave
  vector and neglect couplings between wave vectors that differ by multiples of
  ${\bm q}$. This is sufficient to determine the static properties of the
  helimagnon, but it precludes taking the ferromagnetic limit ${\bm q}\to 0$.
  See also Sec.\ \ref{subsec:IV.D} and Ref.\
  \protect\onlinecite{mode_coupling_footnote}.}
\bibitem[{mod()}]{mode_coupling_footnote}
\bibinfo{note}{This additive procedure is valid as long as ${\bm q}\neq 0$ and
  for wave numbers smaller than $\vert{\bm q}\vert$. As Sec.\
  \ref{subsubsec:IV.C.3} shows, the phase fluctuations correctly turn into the
  ferromagnetic soft modes as ${\bm q}\to 0$, so no ${\bm k}=0$ modes must be
  added in the ferromagnetic limit. See also Ref.\
  \protect\onlinecite{fm_limit_footnote}.}
\end{thebibliography}
%\input{helimagnons_I_rev.bbl}

\end{document}